\documentclass[pre,twocolumn,preprintnumbers,amsmath,amssymb,floatfix,longbibliography,superscriptaddress]{revtex4-2}
\usepackage{graphicx}
\usepackage{epstopdf}
\usepackage{psfrag}
\usepackage{hyperref}
\usepackage{multirow}
\usepackage[sort&compress]{natbib}
\usepackage{amssymb}
\usepackage{amsmath}
\usepackage{amsthm}
\usepackage{bbold}
\usepackage{bbm}
\usepackage{enumerate}
\usepackage{textcomp}
\usepackage{verbatim}
\usepackage{soul}
\usepackage[normalem]{ulem} 
\usepackage{float}
\usepackage{leftindex}

\usepackage{mathtools}
\usepackage[caption=false]{subfig}
\usepackage{orcidlink}
\usepackage{siunitx}
\usepackage{adjustbox}

\DeclareMathOperator{\artanh}{artanh}

\newcommand{\bk}{{\boldsymbol{k}}}

\newcommand{\arctanh}{{\rm artanh}}
\newcommand{\bd}[1]{{\boldsymbol{#1}}}
\newcommand{\beq}{\begin{equation}\begin{aligned}}
\newcommand{\eeq}{\end{aligned}\end{equation}}
\newcommand{\qmaxLn}{\qmax^{\Lambda_n}}
\newcommand{\qmaxLz}{\qmax^{\Lambda}}
\newcommand{\TcLn}{\Tc^{\Lambda_n}}
\newcommand{\tcLn}{\tc^{\Lambda_n}}
\newcommand{\TcL}{\Tc^{\Lambda}}
\newcommand{\tcL}{\tc^{\Lambda}}

\usepackage{xcolor}
\usepackage{cancel}

\definecolor{AC}{rgb}{1, 0.2, 0.7}
\definecolor{SD}{rgb}{1,0,1}

\usepackage[capitalize]{cleveref} 

\crefname{section}{Sec.}{Secs.}
\Crefname{section}{Sec.}{Secs.}
\crefname{appendix}{App.}{Apps.}
\Crefname{appendix}{App.}{Apps.}

\newcommand{\qmax}{q_{\rm max}}
\newcommand{\qbar}{\bar{q}}
\newcommand{\Tc}{T_{\rm c}}
\newcommand{\tc}{t_{\rm c}}

\newcommand{\tauc}[1]{\tau_{\rm c}^\Lambda(#1)}

\usepackage{placeins}

\begin{document}

\hypersetup{pdftitle={title}}
\title{Families of planar lattices with arbitrarily high $\Tc$ for the ferromagnetic Ising model}

\author{Davidson Noby Joseph\,\orcidlink{0009-0008-2420-0951}}
\affiliation{Department of Physics, University of Alberta, Edmonton, Alberta, Canada}
\affiliation{Theoretical Physics Institute, University of Alberta, Edmonton, Alberta, Canada}

\author{Connor M. Walsh\,\orcidlink{0009-0002-6311-9108}}
\affiliation{Department of Physics, University of Alberta, Edmonton, Alberta, Canada}
\affiliation{Theoretical Physics Institute, University of Alberta, Edmonton, Alberta, Canada}

\author{Igor Boettcher\,\orcidlink{0000-0002-1634-4022}}
 \affiliation{Department of Physics, University of Alberta, Edmonton, Alberta, Canada}
 \affiliation{Theoretical Physics Institute, University of Alberta, Edmonton, Alberta, Canada}
 \affiliation{Quantum Horizons Alberta, University of Alberta, Edmonton, Alberta, Canada}

\begin{abstract}
We construct families of periodic tessellations of the plane with arbitrarily high critical temperature, $T_{\rm c}$, for the classical nearest-neighbor uniform ferromagnetic Ising model. Our approach is motivated by recently found exact bounds, which imply that large values of $T_{\rm c}$ require large values of the maximal coordination number of the lattice, $q_{\rm max}$. We create such lattices  through iterative triangulation and derive explicit expressions for their $T_{\rm c}$. Furthermore, we show that $T_{\rm c}$ for these families scales asymptotically as ${\Tc/J\sim A \ln \qmax -2 \ln \ln q_{\rm max}}$ with a universal prefactor ${A=2/\ln 2}$. We introduce a function $\Tc^*(\qmax)$ that we conjecture to be an upper bound on the critical temperature of any periodic tessellation of the plane. We show that the family of so-called Apollonian lattices, which are derived from the Triangular lattice through iterative triangulation, saturates this bound. The lattices discussed in this work are relevant for theoretical questions of optimality in network systems and may be realized experimentally in Coherent Ising Machines or topoelectric circuits in the future.
\end{abstract}
\maketitle

\section{Introduction}

The Ising model is a cornerstone of statistical physics and ubiquitous in the description of critical phenomena \cite{McCoy2012,Fisher1981,Kulske2025}. Since its conception over a century ago to explain magnetism \cite{Ising25}, the model has found many applications in a wide range of fields from biological and social systems, to network theory and machine learning \cite{Alonso2025,Majewski2001,Weber2016,Noble2015,Hopfield1982,Fahlman1983,Ackley1985,Sun2024}. The ferromagnetic Ising model exhibits a phase transition in two and higher dimensions at a finite critical temperature $T_{\rm c}$. Although the critical behavior of correlation functions through the critical exponents is universal, i.e. dependent on dimension but not on the underlying lattice structure, $T_{\rm c}$ is a non-universal quantity determined by the lattice together with the magnetic interaction strength $J_{ij}$ between sites \cite{PRIVMAN1991,vasilyev2019,vasilyev2023}. 

 Recently, the Ising model gained immense interest as an experimental framework for solving various NP-complete optimization problems like MAX-CUT, where the solution of the problem is embedded in the ground state of the model with tunable $J_{ij}$ using so-called Ising machines. Various Ising machines have been realized experimentally with laser systems, initially injection-locked laser systems \cite{Utsunomiya2011,Kenta2012} and later networks of degenerate optical parametric oscillators called Coherent Ising Machines \cite{Wang2013,Marandi2014,Yamamoto2017,Yamamoto2020}. The latter have emerged as a platform for simulating the Ising model on various graph-topologies with tunable couplings and have been used to measure $\Tc$, for instance, in the all-to-all mean field network \cite{Fang2021} and the two-dimensional square lattice \cite{Takesue2023,Inaba2023}. 

For many practical applications of magnetically ordered systems, it is beneficial to investigate which lattices can exhibit high critical temperatures.  If the ferromagnetic interactions are uniform, $J_{ij}=J>0$, large critical temperatures are achieved in high dimensions. Indeed, Fisher and Gaunt showed that for the $d$-dimensional hypercubic lattice we have $\Tc/J \sim 2d$ asymptotically for large $d$ \cite{Fisher1964}. (Here and in the following we set Boltzmann's constant $k_{\rm B}= 1$.)
For two-dimensional periodic tilings by regular polygons, the Triangular lattice with $T_{\rm c}/J =3.641$ features the highest  critical temperature among all $1248$ $k-$uniform tilings with $k\leq 6$ types of vertices \cite{GalebachWebpage,Sanchez1,Sanchez2,SanchezWebpage,GomJau-Hogg,Grunbaum2016-rs,Portillo}.
This finding was recently explained in Ref. \cite{Joseph2026}, where an exact bound on $T_{\rm c}$ for any two-dimensional periodic tiling of the plane was derived. The bound on $T_{\rm c}/J$ is determined by the maximal coordination number of the lattice, $q_{\rm max}$, defined as the largest number of nearest neighbors of any site on the lattice. Since all $k$-uniform lattices with $k\leq 6$ have $q_{\rm max}\leq 6$, their values of $T_{\rm c}/J$ are below the Triangular lattice, which saturates the bound for $q_{\rm max}=6$. However, if  instead of using regular polygons we allow for arbitrary polygons, then lattices with larger $q_{\rm max}$ and higher $T_{\rm c}/J$ can be constructed, such as the Laves-Star (also called Asanoha or hemp-leaf) lattice with $q_{\rm max}=12$ and $\Tc/J=5.007$ \cite{Syozi1972, Codello2010}, or the Compass-Rose lattice with $q_{\rm max}=24$ and $\Tc/J= 6.492$ \cite{Joseph2026}.

Motivated by these examples, it appears natural to ask whether $T_{\rm c}/J$ can be made arbitrarily large for two-dimensional Ising systems on lattices with large $q_{\rm max}$. For other two-dimensional statistical-mechanics systems, applying concepts of entropic order related to the Pomeranchuk effect, it has recently been shown that critical temperatures can be infinite \cite{Entropic1,Entropic2,Entropic3,Entropic4}. For the Ising case, however, the exact bound derived in Ref. \cite{Joseph2026} imply that $T_{\rm c}/J$ is always finite and asymptotically bounded by $T_{\rm c}/J \leq (2/\pi)q_{\rm max}$.  In this work, we show that families of lattices with arbitrarily large values of $T_{\rm c}/J$ can be constructed through the method of iterative triangulation. Their critical temperatures scale asymptotically as
\begin{align}
 T_{\rm c}/J \sim A \ln q_{\rm max}-2\ln\ln q_{\rm max}. \label{Tc_asymptotic_growth}
\end{align}
The logarithmic growth in $q_{\rm max}$, in contrast to the bound which grows linearly in $q_{\rm max}$, indicates that actual values of $T_{\rm c}/J$ seem to fall short of coming close to the exact bound for large $q_{\rm max}$. This is supported by the critical temperatures of the Laves-Star and Compass-Rose lattices quoted above. Here, the coefficient $A=2/\ln 2=2.89$ is \emph{universal} for any family of lattices constructed by iterative triangulation. In particular, this scaling is independent of the starting lattice, and even its periodicity.

\begin{figure*}[htpb]
    \centering
    \includegraphics[scale=0.8]{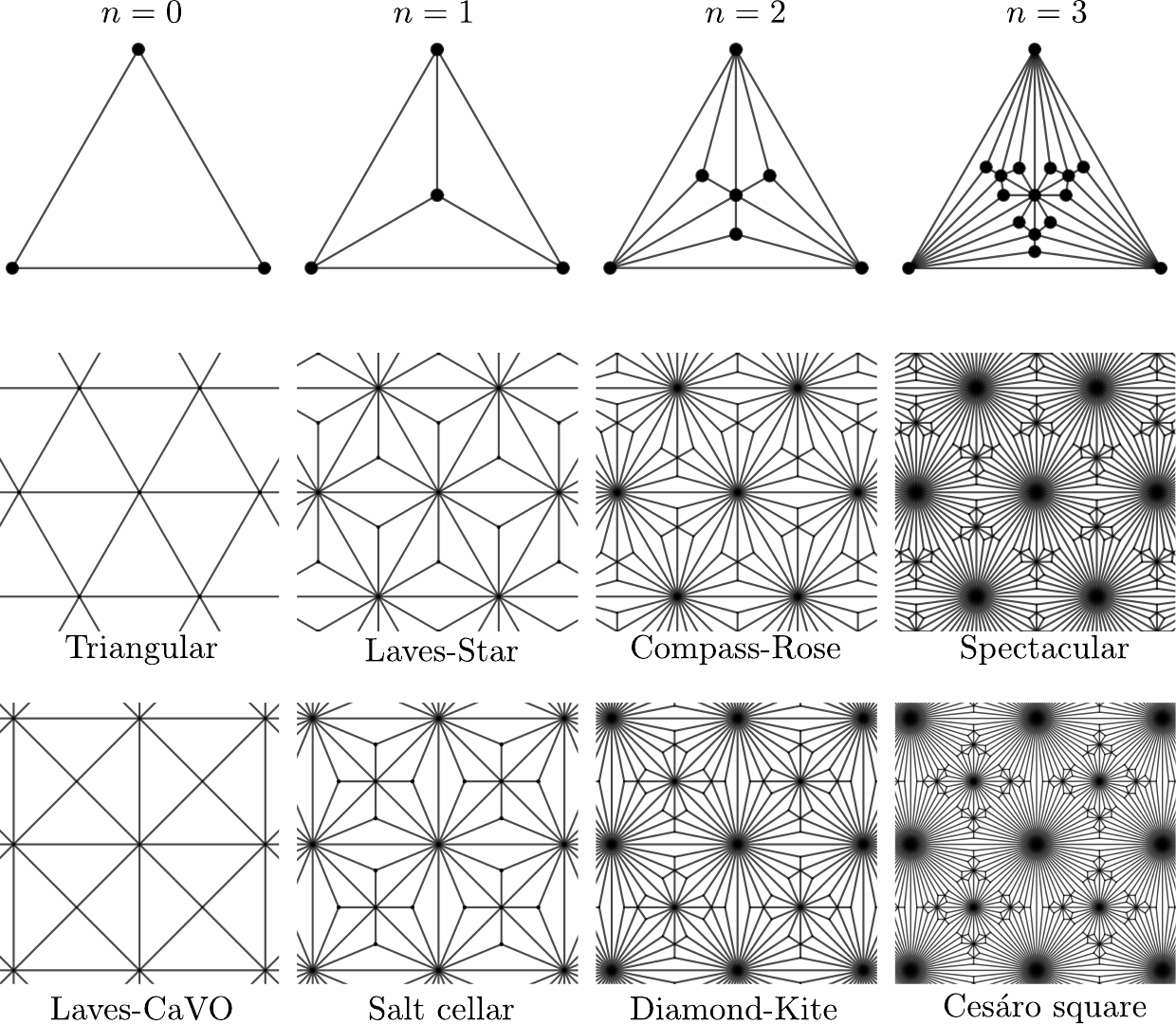}
    \caption{The procedure of iterative triangulation is applied to a single triangle and to two example lattices. The top panel shows the iterative triangulation procedure applied to a generic triangle $n=0,1,2,3$ times. At every step, a new vertex is placed at the center of each existing triangle and connected to its corners. The middle and bottom panels show the procedure applied to the Triangular and Laves-CaVO lattices, producing the Apollonian lattices and the Laves-CaVO family, respectively. For the Apollonian lattices we have the maximal coordination numbers $\qmax=6,12,24,48$, and for the Laves-CaVO family we have $\qmax=8,16,32,64$. With each iteration, $\qmax$ doubles in value, and hence grows exponentially in $n$. }
    \label{ItTrPanel}
\end{figure*}

The outline of this paper is as follows:  in \cref{SecSumMain}, we summarize our main results. In \cref{model}, we introduce the ferromagnetic Ising model and necessary definitions for this work. In \cref{Triangulation}, we define the procedure of iterative triangulation, which we use to generate families of high-$\Tc$ lattices. Next, we derive explicit formulae for the partition function, free energy per site, and critical temperature for lattices under iterative triangulation in \cref{PartitionF}. Then, in \cref{ExampleCalc}, we present the critical temperatures of a collection of base lattices which are periodic tessellations of the plane and their families. We categorize each family in terms of a lattice-dependent constant $K_\Lambda$, which arises from the asymptotic behavior of lattices under iterative triangulation. 
The exact expression for the asymptotic scaling of the critical temperature is then derived in \cref{SecAsym}. Finally, in \cref{comparison}, we investigate $T_{\rm c}$ of various lattice families and define a unique continuous extension $T_{\rm c}^*(q_{\rm max})$ of the critical temperatures of the Apollonian lattices, which we conjecture to be a tight upper bound for the critical temperature of any planar lattice in Euclidean space with $q_{\max}\geq 6$. 

\section{Summary of main results}\label{SecSumMain}

Our construction of high-$T_{\rm c}$ lattices draws inspiration from the Laves-Star and Compass-Rose lattices, which are built from the Triangular and Laves-Star lattices, respectively, by placing a site in the center of each triangle and drawing bonds from it to the sites at the corners of the triangle. This procedure of \emph{iterative triangulation}, which is the main tool of this work, creates an infinite family of lattices, each member with a larger $\qmax$ and higher critical temperature than its predecessor. We illustrate the process on an arbitrary triangle in \cref{ItTrPanel}, and show the first few lattices for two families starting from the Triangular and Laves-CaVO lattices. We call the family containing the Triangular, Laves-Star, and Compass-Rose lattices the \emph{Apollonian lattices}, inspired by their tiles being Apollonian networks \cite{andrade_2005,HeavyTail}. The technique of iterative triangulation can be applied to any planar lattice that is a triangulation, i.e a lattice that consists solely of triangles, to obtain a lattice with larger $\qmax$ and with a critical temperature $T_{\rm c}'/J$ that is higher than that of the original lattice. 

We introduce the temperature weight $t = \tanh(\beta J)$ 
with $\beta=1/T$. The critical weight $t_{\rm c}$ is related to $T_{\rm c}$ by
\begin{align}
 \frac{T_{\rm c}}{J} = \frac{1}{\text{artanh}(t_{\rm c})}.
\end{align} 
We show that after one step of iterative triangulation, the critical weights of the original and triangulated lattices, $t_{\rm c}$ and $t_{\rm c}'$, satisfy
\begin{equation}\label{gforward}
     \tc^\prime=g\left(\tc\right),
\end{equation}
where 
\begin{equation}\label{gdef}
   g(t)=\frac{2t}{1+t+\sqrt{1+6t-7t^2}}.
\end{equation}
We derive this formula using the star-triangle identity \cite{Au1989,Baxter1978}. Similarly, recursive relations between the partition functions and free energy densities follow from
\begin{align}
  \mathcal{Z}^{\prime}(t) & \propto \mathcal{Z}\bigl(h(t)\bigr), \label{partasymp}
\end{align}
with explicitly known $t$-dependent prefactor, and
\begin{equation}\label{h(t)}
    h(t)= g^{-1}(t) = \frac{t(1+t)}{1+t(2t-1)}.
\end{equation}
The full formulae are given in \cref{indz,indf}.

As an illustrative example, consider again the Triangular lattice with $t_{\rm c}=2-\sqrt{3}$.  The Laves-Star lattice in \cref{ItTrPanel} is constructed from the Triangular lattice through iterative triangulation, and we indeed confirm for the Laves-Star lattice that
\begin{equation}
    \frac{\Tc^\prime}{J}=\frac{1}{\arctanh\bigl(g(t_{\rm c})\bigr)}=5.007.
\end{equation}
The next member in the family, obtained from the Laves-Star lattice through iterative triangulation, is the Compass-Rose lattice from  \cref{ItTrPanel}, for which we confirm
\begin{equation}
    \frac{\Tc^{\prime\prime}}{J}=\frac{1}{\arctanh\bigl(g(t_{\rm c}^\prime)\bigr)}=6.492.
\end{equation}
Both critical temperatures, of course, agree with the values quoted in the introduction. Iterating once more, we obtain the Spectacular lattice shown in \cref{ItTrPanel}, with $q_{\rm max}=48$ and
\begin{equation}
    \frac{\Tc^{\prime\prime\prime}}{J}=\frac{1}{\arctanh\bigl(g(t_{\rm c}^{\prime\prime})\bigr)}=8.062.
\end{equation} 
Although our example used the familiar Triangular lattice as a base, we emphasize the remarkable fact that \cref{gforward} is valid for any triangulation. In particular, this equation is valid for triangulations in non-Euclidean space, such as hyperbolic lattices \cite{Wu1996,Wu2000,Krcmar2008,Iharagi2010,Maciejko2020,Boettcher2022,Nussinov25}, if the weight $t_{\rm c}$ of the base lattice is known.

Applying the procedure multiple times to a base triangulation $\Lambda$ with $\tc=\tc^\Lambda$, we  construct a family of lattices $\Lambda_n$, where $n$ refers to the number of iterations performed. For $n\gg1$, the critical temperature $\Tc^{\Lambda_n}$ grows asymptotically as
\begin{equation}\label{asympqmaxn}
        \frac{\Tc^{\Lambda_n}}{J}=2(n-\ln n)+\kappa(\tc^\Lambda) + o(1).
    \end{equation}
Here $\kappa(t)$ is a universal function defined in \cref{kappa}. This highlights the fact that one can tractably attain high-$\Tc$ lattices by applying sufficiently many iterations. 

To compare lattices among different families, and determine which ones have the highest $T_{\rm c}$, we denote the maximal coordination number of the $n^{\rm th}$ iterate $\Lambda_n$ by $q_{\rm max}^{\Lambda_n}$, and by $q_{\rm max}^\Lambda$ the corresponding value for the base lattice $\Lambda=\Lambda_0$. We have 
\begin{align}
 q_{\rm max}^{\Lambda_n} = 2^nq_{\rm max}^\Lambda,
\end{align}
such that \cref{asympqmaxn} implies
\begin{align}\label{asympqmax}
    \frac{T_{\rm c}^{\Lambda_n}}{J} &= A \ln \qmax^{\Lambda_n} - 2\ln \ln \qmax^{\Lambda_n} - K_{\Lambda} + o(1),
\end{align}
with $A=2/\ln 2$ and $K_\Lambda$ a lattice-dependent constant defined in \cref{K_Lamdbda1}. While the leading terms are universal, the subleading corrections contain a constant $K_{\Lambda}$ that depends solely on the base triangulation. As a special case, we have $K_{\Delta}=1.024$ for the Triangular lattice $\Delta$.

\begin{figure*}[tb]
    \centering
    \subfloat[\label{fig_tc_vs_qmax_with_bounds_lin}]{\includegraphics[width=0.5\linewidth]{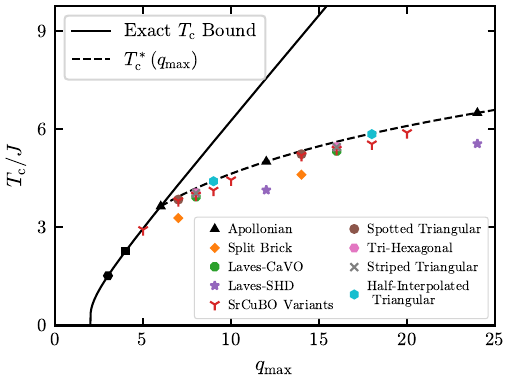}}
    \subfloat[\label{fig_tc_vs_qmax_with_bounds_log}]{\includegraphics[width=0.5\linewidth]{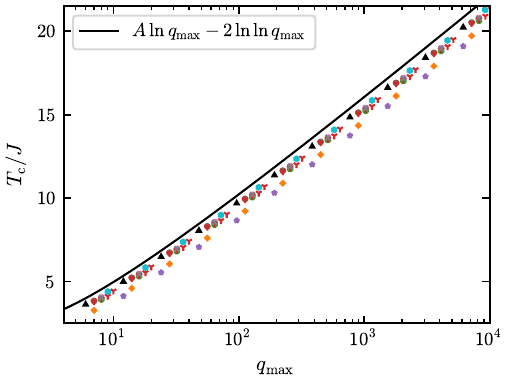}}
    \caption{Plots of lattice critical temperatures $\Tc$ as a function of $\qmax$. Each set of symbols represents a different family of lattices derived from one of the base lattices in \cref{fig_base_lattices}, with the Apollonian lattices denoted by the triangle symbols. (a) The solid line represents the exact analytic upper bound on $\Tc$ given by $ \tanh (J/\Tc)=\tan\bigl(\pi/(2\qmax)\bigr)$~\cite{Joseph2026}, while the dashed line denotes the values of $\Tc^*$, defined as the unique continuation of the critical temperature of the Apollonian lattices as described in \cref{comparison}. All lattices considered have critical temperatures equal to or below $\Tc^*$. Also included here are the Honeycomb lattice (hexagon symbol, $\qmax=3$), Square lattice (square symbol, $\qmax=4$), and SrCuBO lattice (tri-point symbol, $\qmax=5$), which are members of the 1-uniform tilings known as the Archimedean lattices \cite{Farnell2018}. The Honeycomb, Square, and Triangular lattices are the only lattices known to saturate the exact bound shown by the solid curve. (b) Lattice families obtained by $n\leq10$ iterations. Here, the solid line represents the universal asymptotic scaling of the critical temperatures under iterative triangulation given in \cref{Tc_asymptotic_growth}. For each family, $\Tc$ grows linearly in $\ln\qmax$ with a universal slope given by $A=2/\ln2$. The subleading corrections contain a downward shift by a constant $K_\Lambda$ which depends on the base triangulation and is discussed further in \cref{SecAsym}.}
    \label{fig_tc_vs_qmax_with_bounds}
\end{figure*}

For a given value of $q_{\rm max}$, we find that the values of $T_{\rm c}/J$ for all families considered in this work lie below the curve $T_{\rm c}^*(q_{\rm max})/J$ defined through
\begin{equation}
 \frac{\Tc^*(q)}{J} = \frac{1}{\arctanh\bigl(\tc^*(q)\bigr)},  
\end{equation}
where
\begin{equation}\label{tstarmain}
\tc^*(q) = \lim_{n\to\infty} h^n \Bigl( \Bigl[A\ln \bigl(2^nq \bigr) - 2\ln \ln \bigl(2^nq \bigr) - K_\Delta \Bigr]^{-1} \Bigr),
\end{equation}
using the $n^{\rm th}$ functional iterate
\begin{equation}\label{hn}
    h^n=\underbrace{h\circ h\circ \dots \circ h}_{n\text{ times}}.
\end{equation}
The Apollonian lattices satisfy
\begin{align}
    T_{\rm c}^{\Delta_n} &= T_{\rm c}^*(q_{\rm max}^{\Delta_n}). \label{T_star_def}
\end{align}
We conjecture that $T_{\rm c}^*(q_{\rm max})$ is the ultimate upper bound in Euclidean space for $T_{\rm c}$ for all $q_{\rm max}\geq 6$, and therefore replaces the exact bound derived in Ref.~\cite{Joseph2026}, which is asymptotically given by $T_{\rm c}/J\leq(2/\pi)q_{\rm max}$. In \cref{fig_tc_vs_qmax_with_bounds}, we show exact critical temperatures for families of lattices derived from various base triangulations. The exact $\Tc$ bound is shown in \cref{fig_tc_vs_qmax_with_bounds_lin}, 
alongside the curve for $\Tc^*(\qmax)$, which bounds the critical temperatures of all lattices in the plot when $\qmax\geq6$. The universal asymptotic growth of $\Tc$ under iterative triangulation in the regime of large $\qmax$ is shown in \cref{fig_tc_vs_qmax_with_bounds_log}.

\section{Ferromagnetic Ising Model}\label{model}
We consider the ferromagnetic Ising model on a two-dimensional planar lattice with classical spin variables $s_i=\pm1$ at sites $i$ of the lattice and uniform ferromagnetic exchange energy or coupling $J>0$. The Hamiltonian for the system is given by
\begin{equation}\label{Ham}
    H=-J \sum_{\langle i,j\rangle } s_i s_j,
\end{equation}
where the sum is over all nearest-neighbors on the lattice. The lattice can be interpreted as a graph where the vertices correspond to the sites, the edges correspond to the bonds of the lattice, and the faces correspond to the closed polygons that tile the plane. Denote by $\mathcal{V}$ the set of vertices, $\mathcal{E}$ the set of edges, and $\mathcal{F}$ the set of faces. Through this graph-theoretic framework, we identify the total number of vertices $V$, edges $E$, and faces $F$ corresponding to the total number of sites, bonds, and tiles on the lattice with $V=|\mathcal{V}|,\ E=|\mathcal{E}|$, and $F=|\mathcal{F}|$. For the sum in \cref{Ham}, $\langle i,j\rangle=\langle j, i\rangle  \in\mathcal{E}$ corresponds to the edge between vertices $i$ and $j$.

On immersing the system in a thermal bath at temperature $T$, the partition function reads
\begin{equation}\label{part1}
    \mathcal{Z}= \sum_{\{s_\ell\}}e^{-\beta H},
\end{equation}
where the sum  is over all spin configurations $\{s_\ell\}_{\ell\in \mathcal{V}}$ and $\beta=1/T$ is the inverse temperature. For our work, we express the partition function in terms of the temperature variable 
\begin{equation}
    t=\tanh(\beta J)\in (0,1).
\end{equation}
We apply periodic boundary conditions and consider the partition function for a finite graph on a torus. Using the van der Waerden identity \cite{van1941} $e^{\beta J s_is_j}=\cosh(\beta J)+ s_is_j \sinh(\beta J)$ and expanding the exponential in \cref{part1} as a product over the nearest-neighbors, we arrive at the exact expression for the partition function \cite{cimasoni2012}
\begin{equation}\label{part2}
    \mathcal{Z}(t)= (1-t^2)^{-E/2}\sum_{\{s_\ell\}} \prod_{\langle i,j\rangle\in \mathcal{E} } (1+t s_i s_j ),
\end{equation}
valid for all $t$. In view of \cref{part2}, $t$ can now be interpreted as a weight attached to each edge $\langle i,j\rangle$ connecting the vertices $i$ and $j$ on the finite graph. Thus, throughout this work, we shall refer to $t$ as a \emph{weight} on the graph. For infinite lattices, the expression becomes the high-temperature expansion for small $t$. In this work, we derive all results for finite graphs and only take the thermodynamic limit at the end.

The free energy per site is defined using the partition function as
\begin{equation}\label{f_def}
    -\beta f(t)= \lim_{V\to\infty}\frac{\ln\mathcal{Z}(t)}{V}.
\end{equation}
The critical weight $\tc=\tanh(J/\Tc)$ is defined as a non-analytic point of the free energy per site in the thermodynamic limit, which can be used to deduce the critical temperature through
\begin{equation}\label{tctanh}
    \frac{\Tc}{J}=\frac{1}{\arctanh(\tc)}.
\end{equation}
For planar lattices, the exact analytical expression for $f(t)$ can be readily derived through the Kac--Ward {formalism} in terms of the Kac--Ward matrix $W$ \cite{PhysRev.88.1332,Feynman}. If the lattice is periodic, then $f(t)$ can be expressed as an {integral} over the Brillouin Zone involving the momentum-space Kac--Ward matrix $W(\bk)$ \cite{Kardar,IsingArch,Joseph2025}. This can be used to calculate the critical weight by solving the equation
\begin{equation}
   \label{KacWardTc} \det\bigl(\mathbbm{1}-\tc W(\bd{0})\bigr)=0.
\end{equation}
For planar lattices that are not periodic, $t_{\rm c}$ does not follow from such a simple result, but is defined as a non-analytic point in the free energy per site, $f(t)$, in the thermodynamic limit.

\section{Lattices under Iterative Triangulation}\label{Triangulation}

A planar lattice can be viewed as a set of polygons which fully tile the plane. We consider here the special case where the only polygons used are triangles, in which case the tiling is called a \textit{triangulation}. The number of edges leaving a vertex $i$ is denoted by the coordination number $q_i$. The average coordination number is given by
\begin{equation}\label{qbardef}
    \bar{q}=\frac{1}{V}\sum_{i=1}^{V} q_i = \frac{2E}{V},
\end{equation}
where we used the fact that summing over all $q_i$ involves counting each edge twice.

\begin{figure}[tb]
    \centering
    \subfloat[\label{fig_iterated_triangulation_a}]{\includegraphics[width=0.65\linewidth]{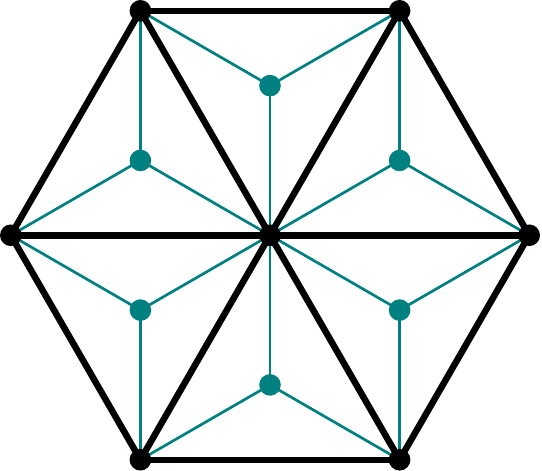}} \\
    \subfloat[\label{fig_iterated_triangulation_b}]{\includegraphics[width=0.65\linewidth]{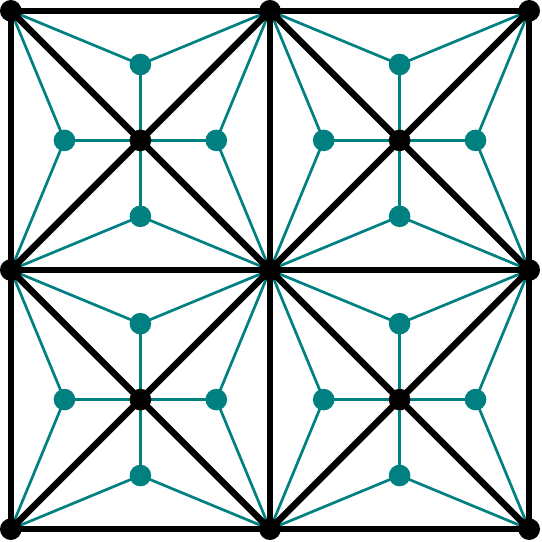}}
    \caption{Examples of the iterative triangulation procedure applied to the Triangular lattice (top) and the Laves-CaVO lattice (bottom). Base lattices are outlined by the thick black lines. New vertices introduced by the triangulation procedure are represented by the teal points, with the new bonds denoted by thin teal lines. In each case, iterative triangulation keeps the average coordination number $\qbar$ fixed at 6 but doubles the maximal coordination number $\qmax$, here from $6$ to $12$ (top) and from $8$ to $16$ (bottom).}
    \label{fig_iterated_triangulation}
\end{figure}

We begin with an arbitrary triangulation of the plane, which we will call the \emph{base lattice}, denoted by $\Lambda$. 
The iterative triangulation procedure is performed by placing one new vertex inside each triangular face of the base lattice and connecting each new vertex to the three vertices of the face containing it. This produces a new lattice where each face of the base lattice has been subdivided into three new triangular faces. In \cref{fig_iterated_triangulation}, we show this procedure applied to the Triangular (top panel) and Laves-CaVO (bottom panel) lattices. The iterative procedure can be applied again to the resultant lattice to obtain yet another lattice. We denote by $\Lambda_n$ the lattice resulting from applying this procedure $n$ times, where the base lattice corresponds to $n=0$. Explicit lattices for $n\leq3$ are shown in \cref{ItTrPanel}.

We highlight a few geometric facts that will be useful for deriving the partition function of the resultant lattice. First, note that any triangulation has an average coordination number $\bar{q}=6$. This result can be readily deduced from $V - E + F = \chi$, where $\chi$ is the Euler characteristic. For finite graphs with periodic boundary conditions (topologically a torus with $\chi=0$), we thus have
\begin{equation}\label{Euler_characteristic}
    F = E - V. 
\end{equation}
If the tiling is a triangulation, then each face is bounded by three edges, and each edge is shared by two triangles, which implies $2E=3F$. Together with \cref{Euler_characteristic}, this gives
\begin{align}
     F&=2V,\label{F_2V}\\
     E&=3V, \label{E_3V} 
\end{align}
and thus
\begin{equation}\label{qbar_6}
    \qbar=\frac{2E}{V} = 6.
\end{equation}
Therefore, since iterative triangulations are triangulations themselves, the average coordination number $\bar{q}$ of the lattice is preserved.

The maximal coordination number $\qmax^\Lambda$ of the base lattice $\Lambda$, however, is successively doubled under iterative triangulation. Each site $i$ on the base lattice is surrounded by $q_i$ faces, and is therefore connected to $q_i$ new vertices under triangulation. In particular, any site with $\qmax^\Lambda$ neighbours on the base lattice will have $2\qmax^\Lambda$ neighbours in the resultant lattice. If we define the maximal coordination number of the lattice $\Lambda_n$ by $\qmax^{\Lambda_n}$, we therefore have
\begin{align}
    \qmax^{\Lambda_n} &= 2^n\qmax^\Lambda.\label{qmaxn}
\end{align}

Finally, we note that iterative triangulation triples the number of vertices, edges, and faces of the base lattice. We denote by $F'$, $E'$, and $V'$ the number of faces, edges, and vertices on the resultant lattice, respectively. The procedure divides each face into three triangles, so we have $F'=3F$. It also introduces one new vertex into each face of the base lattice yielding
\begin{equation}\label{3V}
    V' = V + F = 3V
\end{equation}
by \cref{F_2V}. 
Finally, as the new lattice is itself a triangulation, \cref{E_3V} also gives
\begin{equation}\label{3E}
    E' = 3V' = 3E.
\end{equation}
Applying these relations successively, we find that the lattice resulting from $n$ iterations has
\begin{align}
    F_n &= 3^n F, \label{Fn} \\
    V_n &= 3^n V, \label{Vn} \\
    E_n &= 3^n E, \label{En} \\
    \qbar_{n} &= \qbar = 6, \label{qbarn}
\end{align}
where $F_n$, $V_n$, and $E_n$ are the number of faces, vertices, and edges, respectively, of the lattice $\Lambda_n$, and $\qbar_n$ its average coordination number.

\section{Partition function and free energy per site under triangulation}\label{PartitionF}
Now that we have described the various families of lattices, we proceed to describe their exact partition function under iterative triangulation. The formalism can be applied on top of any base lattice $\Lambda$ that is a triangulation. For illustrative purposes, let us consider the Triangular lattice $\Delta$ as our base lattice, i.e $\Lambda=\Delta$. We call its corresponding family under iterative triangulation the \emph{Apollonian lattices}, denoted by $\Lambda_n=\Delta_n$ for $n\geq 1$, where $\Lambda_1$ is the Laves-Star lattice depicted in \cref{fig_iterated_triangulation_a}. 

The partition function $\mathcal{Z}_1$ of the Laves-Star lattice on a finite lattice with periodic boundary conditions is found from \cref{part2} as
\begin{equation}
    \mathcal{Z}_{1}(t)= (1-t^2)^{-E_1/2}\sum_{\{s_\ell\}} \prod_{\langle i,j\rangle\in \mathcal{E}_1 } (1+ ts _i s_j ),
\end{equation}
where $\mathcal{E}_1$ contains the edges of the Laves-Star lattice and $E_1=|\mathcal{E}_1|$ is its size. Since the Laves-Star lattice is constructed through iterative triangulation, the spins on the vertices of this lattice $\{s_{\ell}\}$ can be partitioned into the spins $\{s_m\}$ on the underlying Triangular lattice, each with six edges emanating from it, and the spins $\{s^{\prime}_a\}$ at the center of each triangle, each with three edges connected to it. For the Laves-Star lattice, this partitioning is illustrated in \cref{fig_iterated_triangulation_a}, where the spins $\{s_m\}$ correspond to the sites on the base lattice (shown in black), while the spins $\{s^{\prime}_a\}$ correspond to the new sites (shown in teal). In this manner, we split the sum over all the spin configurations into the sum over the spin configurations on the vertices of the Triangular lattice and over the spins at its center. Importantly, this allows us to split the product over all the edges in $\mathcal{E}_1$ into those over the finite Triangular lattice $\mathcal{E}_0$ with periodic boundary conditions of size $E_0=|\mathcal{E}_0|$, and the left-over edges $\mathcal{E}_{1}\backslash\mathcal{E}_{0}$ that arise in the construction as 
\begin{align}\label{partsum1}
    \nonumber \mathcal{Z}_{1}(t)&= (1-t^2)^{-E_1/2} \sum_{\{s_m\}} \left[\prod_{\langle i, j\rangle\in \mathcal{E}_{0}}(1+t s_i s_j)\right]\\& \phantom{\ (1-t^2)^{-E_1/2} }\times \sum_{\{s^{\prime}_a\}}\left[\prod_{\langle k, a\rangle \in \mathcal{E}_{1}\backslash \mathcal{E}_0}(1+t s_k s^{\prime}_a)\right].
\end{align}
Here, the product over $\mathcal{E}_{0}$ is factored out, as it is independent of the sum over $\{s_a^{\prime}\}$, in anticipation of decimating the spins over the centers of all triangles. Explicitly, in \cref{partsum1}, the indices $i$, $j$, and $k$ run over sites on the underlying Triangular lattice while the index $a$ runs over the new sites introduced in the triangulation.

\begin{figure}[tb]
    \centering
    \includegraphics[width=\linewidth]{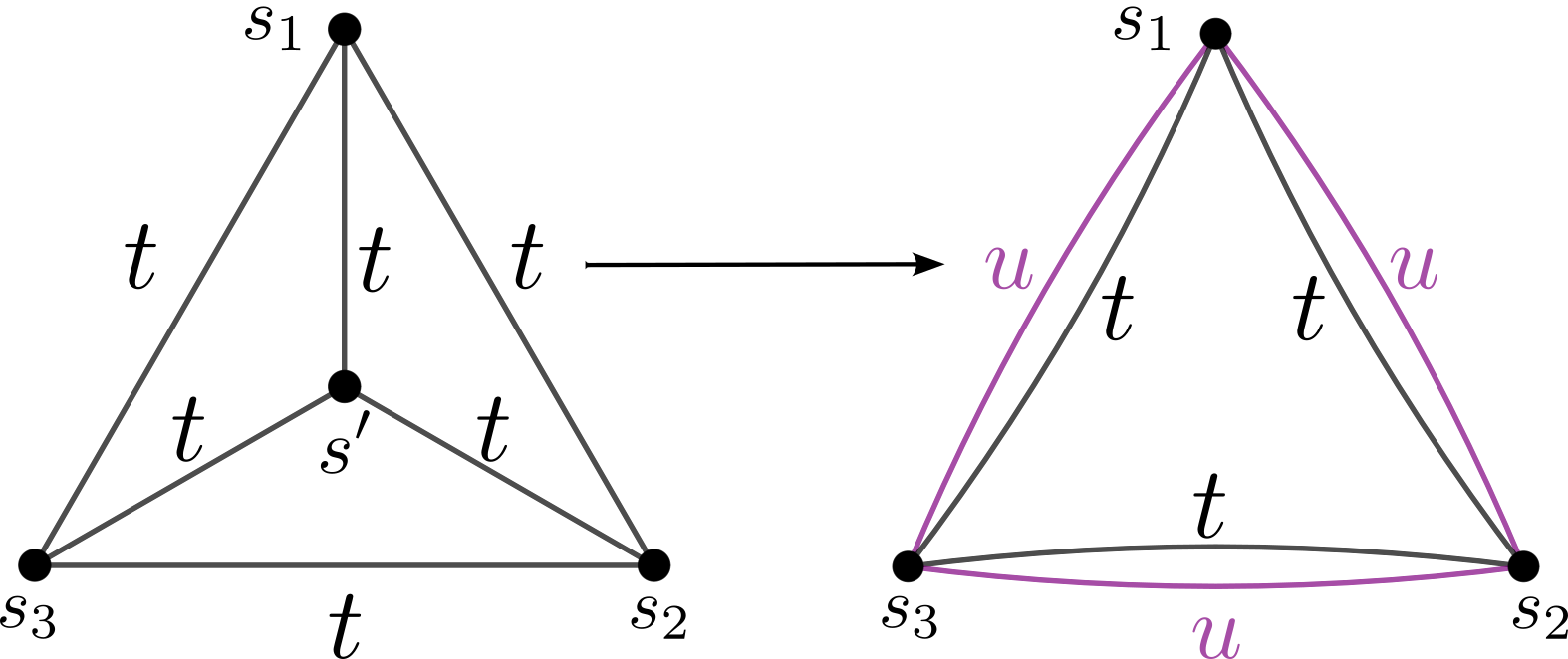}
    \caption{The decimation of the spin $s^\prime$ present inside the star at the center of a triangle produces an effective coupling $u$ per edge in addition to the existing coupling $t$. After decimation, the partition function accounts for each edge with the term $(1+t s_is_j)(1+u s_is_j)^2$ due to the fact that the new contribution $(1+u s_is_j)$ has to be counted twice between two adjacent triangles.}
    \label{STrlattice}
\end{figure}

To perform this decimation, it is sufficient to work with an arbitrary triangle and decimate the spin inside, as the procedure is identical for all triangles. Consider such a triangle with spins $s_1,s_2,s_3$ at its vertices, all connected to a spin $s^{\prime}$ at the center. Decimating the spin $s^\prime$, using the star-triangle identity \cite{Au1989,Baxter1978}, produces
\begin{multline}\label{decim}
    \sum_{s^{\prime}=\pm 1} (1+t s_1 s^{\prime}) (1+t s_2 s^{\prime}) (1+t s_3s^{\prime}) \\ =2 \bigl(1+t^2(s_1s_2+s_2s_3+s_3s_1) \bigr).
\end{multline}
In anticipation of our result, we seek to express \cref{decim} in terms of the product $(1+u s_1 s_2) (1+u s_2s_3) (1+u s_3 s_1)$ with effective weight $u$ and a prefactor. The decomposition

\begin{multline} \label{uGeq}
   2 \bigl(1+t^2(s_1s_2+s_2s_3+s_3s_1) \bigr) \\
   = G^{\frac{3}{2}} (1+u s_1s_2)(1+u s_2s_3)(1+us_3s_1)
\end{multline}
admits two solutions for $u$ and $G$ for all combinations of spins $s_i=\pm1 $, given  by
\begin{align}
    u_{\pm}(t)&=\frac{1+t^2\pm\sqrt{1+2 t^2-3 t^4}}{2 t^2},\label{u(t)}\\
    G(t)&=\sqrt[3]{16^2(3t^2+1)^2}\left(\frac{t^2}{1+3 t^2-\sqrt{1+2 t^2-3 t^4}}\right)^2,\label{Geq}
\end{align}
where the exponent $3/2$ of the prefactor $G$ is chosen for convenience. We discard the positive branch of the square root in \cref{u(t)} as it produces an unphysical divergence of $u(t)$ in the limit of negligible coupling $J$. Hence, we select
\begin{equation}
    \label{ueq}
    u(t)=\frac{1+t^2-\sqrt{1+2 t^2-3 t^4}}{2 t^2}.
\end{equation}
For more details regarding the use of the star-triangle identity and the derivation of \cref{Geq,ueq}, we refer to \cref{sttr}. We conclude that decimating a spin at the center of a generic triangle ($i,j,k$) produces a factor of $G^{\frac{3}{2}} (1+u s_is_j)(1+u s_js_k)(1+u s_ks_i)$ in the partition function, in addition to the existing term $(1+ts_1s_2)(1+ts_2s_3)(1+t s_3s_1)$, as illustrated in \cref{STrlattice}. 

Since edges on the Triangular lattice are shared between two triangles, each edge inherits the weight $u$ twice upon decimating all the new spins, and thus each factor of the form $(1+u s_i s_j)$ appears twice in the partition function. Similarly, $G^{\frac{3}{2}}$ appears once per triangle, giving rise to a factor of $G^{\frac{3}{2}F_0}$, where $F_0$ is the number of faces on the Triangular lattice. 
These factors combine in the partition function to produce
\begin{align}
    \nonumber \mathcal{Z}_{1}(t)={}& \left(\sqrt{1-t^2}\right)^{-E_1} G^{\frac{3}{2}F_{0}} \notag \\&\times\sum_{\{s_m\}} \left[\prod_{\langle i, j\rangle\in \mathcal{E}_{0}}(1+t s_i s_j)(1+u s_i s_j)^2 \right].
\end{align}

Next, we use \cref{ueq} to express the product $(1+t s_is_j)(1+u s_is_j)^2$ as \begin{equation}
    (1+t s_is_j)(1+u s_is_j)^2=H(t) \bigl(1+ h(t) s_is_j\bigr),
\end{equation}
where 
\begin{equation}
    H(t)=\frac{\left(1+t^2+2 t^3\right) \left(1+t^2-\sqrt{1+2 t^2-3 t^4}\right)}{2 t^4}
\end{equation}
is an analytic function on $t\in (0,1)$, and $h(t)$ is given by \cref{h(t)} as
\begin{equation}
    h(t)=\frac{t(1+t)}{1+t(2t-1)}.
\end{equation}
This transformation allows us to rewrite the partition function for the Laves-Star lattice in terms of the partition function of the underlying Triangular lattice with an effective weight $h(t)$. Since each edge appears with a factor of $H$, we arrive at 
\begin{multline}\label{3V1-V1}
    \mathcal{Z}_1(t)=\left(\sqrt{1-t^2}\right)^{-3 V_1} G^{V_1}H^{E_0} \\ \times \left(\sqrt{1-h(t)^2}\right)^{E_{0}}\mathcal{Z}_0\bigl(h(t)\bigr),
\end{multline}
where $V_1$ is the total number of vertices in Laves-Star and $\mathcal{Z}_0$ is the partition function of the Triangular lattice, which satisfies 
\begin{equation}
    \left(\sqrt{1-h^2}\right)^{E_0}\mathcal{Z}_0(h)=\sum_{\{s_m\}} \prod_{\langle i,j\rangle\in \mathcal{E}_0 } (1+ hs _i s_j )
\end{equation}
according to \cref{part2}. The exponent $3V_1$ of the first term in \cref{3V1-V1} is obtained through \cref{E_3V}, whereas the exponent accompanying $G$ is obtained through \cref{F_2V,3V}. Finally using \cref{3E}, we rewrite $E_0$ in terms of $V_1$ to express the partition function of the Laves-Star lattice as
\begin{equation}\label{forwardsZ}
    \mathcal{Z}_1(t)= \left(\frac{H(t)G(t)}{1-t^2}\right)^{V_1}\left( 
    \frac{1-h(t)^2}{1-t^2}\right)^{V_{1}/2} \mathcal{Z}_{0}\bigl(h(t)\bigr).
\end{equation}
This result is true for arbitrary $V_1$, denoting the number of vertices of the finite Laves-Star lattice with periodic boundary conditions. Taking the natural logarithm on both sides, we derive the free energy per site
\begin{equation}
    -\beta f_1(t)=\ln\biggl(\frac{H(t)G(t)}{1-t^2}\biggr)+\frac{1}{2}\ln\biggl(\frac{1-h(t)^2}{1-t^2}\biggr)-\beta f_0\bigl(h(t)\bigr),
\end{equation}
which is true even in the thermodynamic limit. 
Since the free energy density of the Triangular lattice, $f_0(t)$, is non-analytic at $t=\tc^{\Lambda}$, it follows that $f_1(t)$ is critical when $h(t)=\tc^{\Lambda}$. Thus, the effective weight at criticality satisfies the equation
\begin{equation}\label{hLCTr}
h(\tc^{\Lambda_1})=\tc^{\Lambda},
\end{equation}
where $\tc^{\Lambda}=2-\sqrt{3}$ is the critical weight of the Triangular lattice. 
We invert \cref{hLCTr} to obtain
\begin{equation}
\tc^{\Lambda_1}=g(\tc^{\Lambda}),
\end{equation}
where we recall from \cref{gdef} that 
\begin{equation}
    g(t)=\frac{2t}{1+t+\sqrt{1+6t-7t^2}}.
\end{equation}

Furthermore, we can perform induction to obtain the general expression for the partition function of any Apollonian lattice $\Lambda_n$, using \cref{forwardsZ} as our base case to derive
\begin{align}
    \nonumber \mathcal{Z}_{\Lambda_n}(t)=\mathcal{Z}_{\Lambda}\bigl(h^n(t)\bigr) \label{indz}\prod_{p=0}^{n-1}&\Biggl[\left(\frac{H\bigl(h^p(t)\bigr)G\bigl(h^p(t)\bigr)}{1- h^p(t)^2}\right)^{{V_{n}}/{3^p}} \\&\times \left(\frac{1-h^{p+1}(t)^2}{1-h^{p}(t)^2}\right)^{\frac{1}{2}{V_{n}}/{3^p}}\Biggr].
\end{align}
Here $\mathcal{Z}_{\Lambda}$ is the partition function of the base lattice, $V_{n}$ is the number of vertices in $\Lambda_{n}$, and $h^{p}(t)$ denotes the $p^{\rm th}$ functional iterate of $h$ as defined in \cref{hn}. From this, we recall \cref{f_def} to deduce the free energy per site as
\begin{align}
   \nonumber  -\beta f_{\Lambda_n}(t)=&-\frac{1}{3^n}\beta f_{\Lambda}\bigl(h^{n}(t)\bigr)\\\nonumber &+\sum_{p=0}^{n-1}\frac{1}{3^p}\ln \left(\frac{H\bigl(h^p(t)\bigr)G\bigl(h^p(t)\bigr)}{1-h^p(t)^2}\right)\\&+\frac{1}{2}\sum_{p=0}^{n-1}\frac{1}{3^{p}}\ln\left(\frac{1-h^{p+1}(t)^2}{1-h^{p}(t)^2}\right).\label{indf}
\end{align}
In a similar fashion, the critical weight $\tcLn$ satisfies 
\begin{equation}\label{indhtc}
    h^n(\tcLn)= \tc^\Lambda
\end{equation}
or, equivalently,
\begin{equation}\label{indtc}
    \tc^{\Lambda_n}=g^n(\tc^\Lambda),
\end{equation}
using the $n^{\rm th}$ functional iterate of $g$ or $h$ from \cref{h(t),gdef}. We deduce the critical temperature $\Tc^{\Lambda_n}$ of the lattice $\Lambda_n$ using \cref{tctanh} as
\begin{equation}\label{Tcn}
    \frac{\Tc^{\Lambda_{n}}}{J}= \frac{1}{\arctanh\bigl(g^n(\tc^\Lambda)\bigr)}.
\end{equation}
Although the above derivation used the Apollonian lattices as an example, the procedure is completely general and \cref{indz,indf,indhtc,indtc,Tcn} are valid for any triangulation $\Lambda$ and its family of lattices $\Lambda_n$, since the decimation is confined within a triangle. The derivation of \cref{indz,indf,indhtc,indtc} using induction are presented in \cref{induction}.

\section{Example Calculations for Various Base Lattices} \label{ExampleCalc}

\begin{figure*}[h]
    \centering
    \begin{adjustbox}{margin=0cm 0cm 0.1cm 0cm} 
        \subfloat[Triangular ($\qmax=6$) \label{fig_base_lattices_Triangular}]{\includegraphics[width=0.23\linewidth]{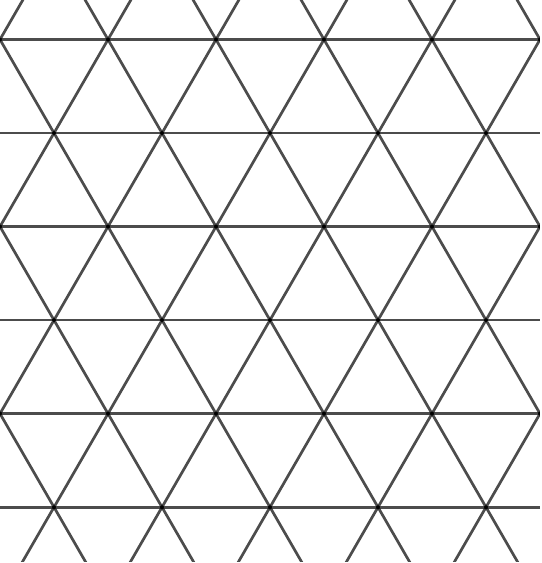}}
    \end{adjustbox}
 \begin{adjustbox}{margin=0.1cm 0cm 0.1cm 0cm} 
        \subfloat[Split Brick ($\qmax=7$) \label{fig_base_lattices_Split_Brick}]{\includegraphics[width=0.23\linewidth]{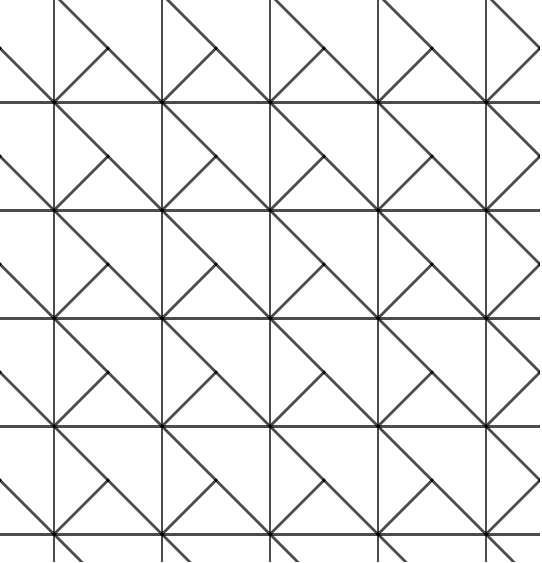}}
    \end{adjustbox}
    \begin{adjustbox}{margin=0.1cm 0cm 0.1cm 0cm} 
        \subfloat[Laves-CaVO ($\qmax=8$) \label{fig_base_lattices_Laves_CaVO}]
        {\includegraphics[width=0.23\linewidth]{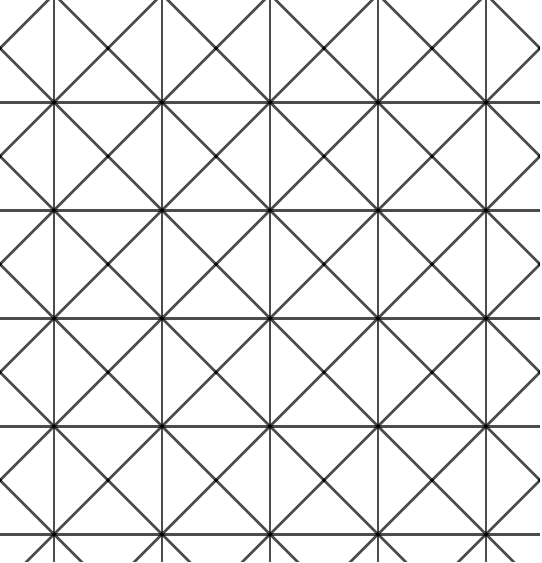}}
    \end{adjustbox}    
    \begin{adjustbox}{margin=0.1cm 0cm 0cm 0cm} 
        \subfloat[Laves-SHD ($\qmax=12$) \label{fig_base_lattices_Laves_SHD}]
        {\includegraphics[width=0.23\linewidth]{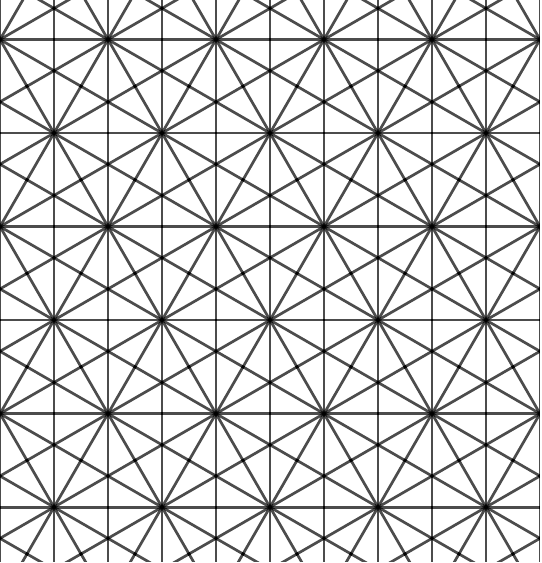}}
    \end{adjustbox}  \\
    \begin{adjustbox}{margin=0cm 0cm 0.1cm 0cm} 
        \subfloat[SrCuBO--7 ($\qmax=7$) \label{fig_base_lattices_SrCuBO7}]{\includegraphics[width=0.23\linewidth]{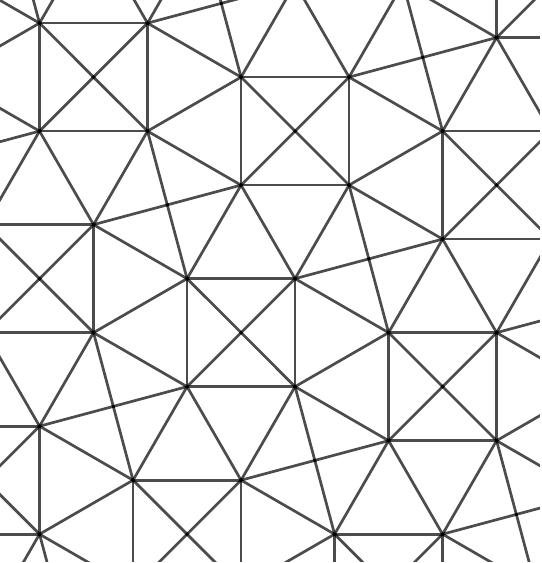}}
    \end{adjustbox}
    \begin{adjustbox}{margin=0.1cm 0cm 0.1cm 0cm} 
        \subfloat[SrCuBO--8 ($\qmax=8$) \label{fig_base_lattices_SrCuBO8}]{\includegraphics[width=0.23\linewidth]{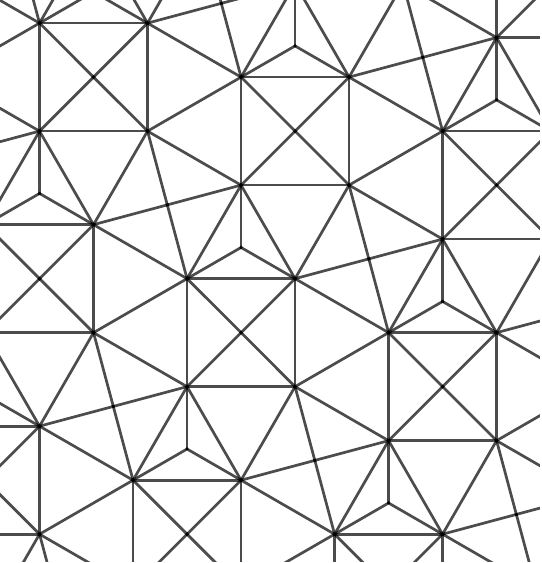}}
    \end{adjustbox}
    \begin{adjustbox}{margin=0.1cm 0cm 0.1cm 0cm} 
        \subfloat[SrCuBO--9 ($\qmax=9$) \label{fig_base_lattices_SrCuBO9}]{\includegraphics[width=0.23\linewidth]{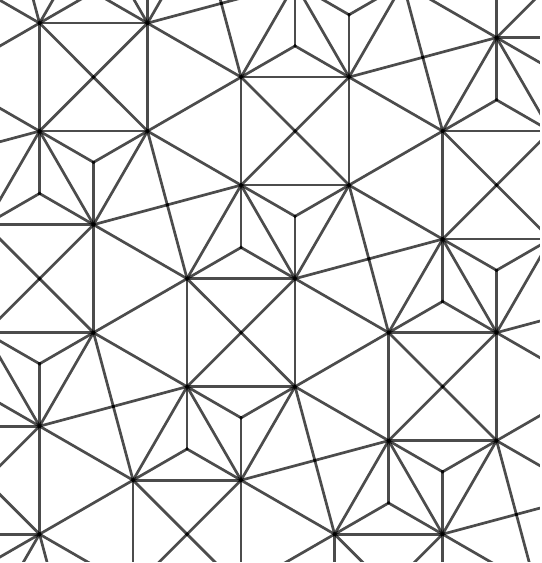}} 
    \end{adjustbox}
    \begin{adjustbox}{margin=0.1cm 0cm 0cm 0cm} 
        \subfloat[SrCuBO--10 ($\qmax=10$) \label{fig_base_lattices_SrCuBO10}]{\includegraphics[width=0.23\linewidth]{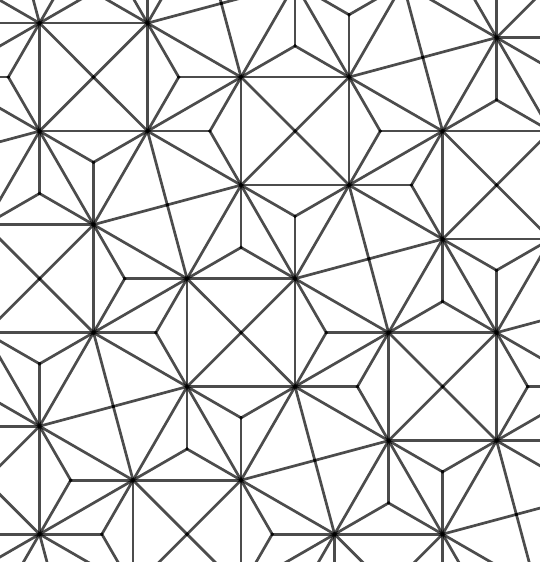}}
    \end{adjustbox} \\
    \begin{adjustbox}{margin=0cm 0cm 0.1cm 0cm} 
        \subfloat[Spotted Triangular ($\qmax=7$) \label{fig_base_lattices_Spotted_Triangular}]{\includegraphics[width=0.23\linewidth]{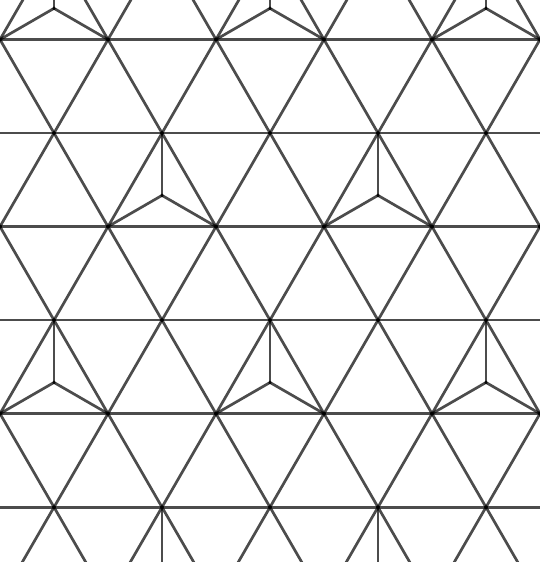}}
    \end{adjustbox} 
    \begin{adjustbox}{margin=0.1cm 0cm 0.1cm 0cm} 
        \subfloat[Tri-Hexagaonal ($\qmax=8$) \label{fig_base_lattices_Tri_Hexagonal}]{\includegraphics[width=0.23\linewidth]{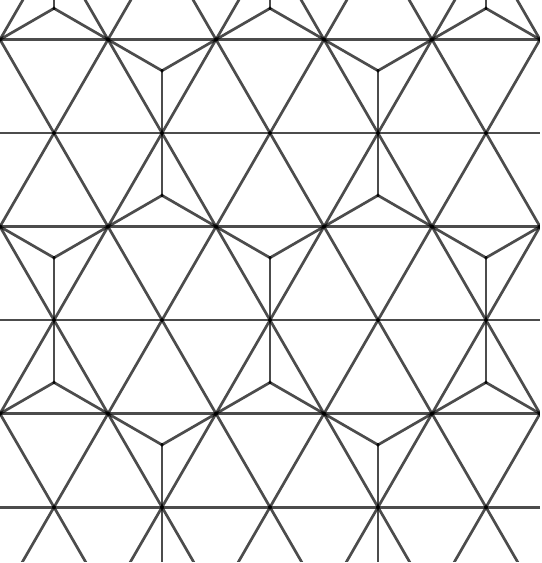}}
    \end{adjustbox}
    \begin{adjustbox}{margin=0.1cm 0cm 0.1cm 0cm} 
        \subfloat[Striped Triangular ($\qmax=8$) \label{fig_base_lattices_Striped_Triangular}]{\includegraphics[width=0.23\linewidth]{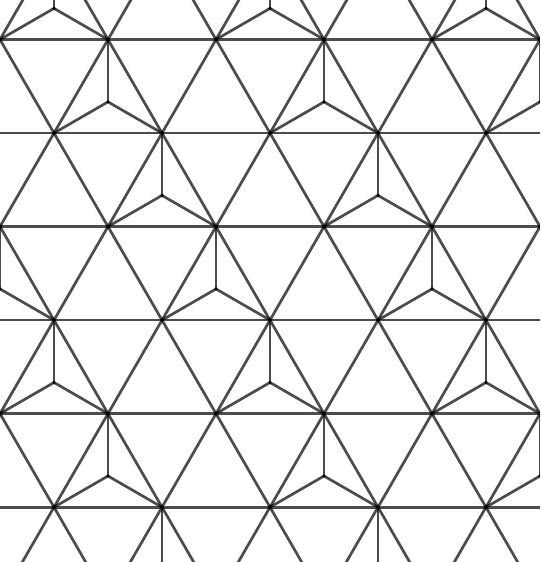}}
    \end{adjustbox}
    \begin{adjustbox}{margin=0.1cm 0cm 0cm 0cm} 
        \subfloat[Half-Interpolated Triangular ($\qmax=9$) \label{fig_base_lattices_Half_Interpolated_Triangular}]{\includegraphics[width=0.23\linewidth]{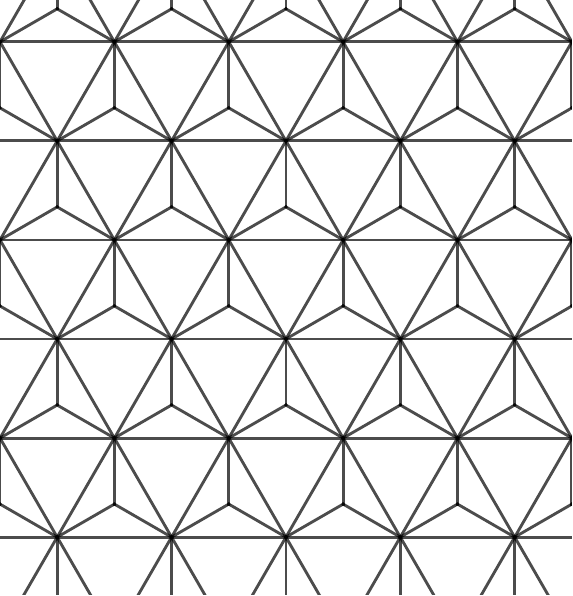}}
    \end{adjustbox}
    \caption{The twelve base lattices with varying maximal coordination number $\qmax$ that are considered in this work are shown. Each base lattice is a triangulation. In the top row, we have the familiar Triangular lattice, as well as the duals of the CaVO and SHD lattices. The Split-Brick lattice with $\qmax=7$ is constructed from the Laves-CaVO lattice by removing one of the bonds from the unit cell. Lattices in the central row are constructed by inserting extra bonds and vertices onto the so-called SrCuBO lattice, which is topologically equivalent to the Shastry--Sutherland lattice \cite{SRIRAMSHASTRY19811069}. Lattices in the bottom row are similarly constructed from the Triangular lattice. Critical temperatures for each of these lattices and their families under iterative triangulation are plotted in \cref{fig_tc_vs_qmax_with_bounds} and are also tabulated in  \cref{tab_base_lat_data}, in increasing order of $K_{\Lambda}$.}
    \label{fig_base_lattices}
\end{figure*}

In this section, we present critical temperatures for various lattice families, each derived from a different base lattice $\Lambda$. The twelve base lattices used in this work, which are depicted in \cref{fig_base_lattices}, are triangulations with diverse values of $\qmax^\Lambda$. Among these, the Triangular, Laves-CaVO, and Laves-SHD lattices are drawn from a collection known as the Archimedean lattices and their duals, the Laves lattices \cite{Joseph2025,Grunbaum2016-rs,Laves1931}. Each of the other nine base lattices considered is a modification of an Archimedean lattice. The critical temperatures for these base lattices are derived using the Kac--Ward formalism described in \cref{model}. For explicit examples of the Kac--Ward method applied to various periodic tessellations, see Ref.~\cite{Joseph2026}.

Once the critical temperature $\TcL$ (and therefore $\tcL$) of a base lattice $\Lambda$ is known, the values of $\TcLn$ for all other lattices in the corresponding family can be readily computed using \cref{Tcn}. The critical temperatures for the Apollonian lattices and for the Laves-CaVO family are collected in \cref{tab_Tn} for up to $n=10$ iterations. \Cref{tab_base_lat_data} shows the critical temperatures for all twelve base lattices, along with the lattice-dependent constants $K_\Lambda$ from \cref{asympqmax}, which differentiate the various families. Base lattices with smaller $K_{\Lambda}$ yield families of lattices whose critical temperatures are larger as a function of $\qmax$.

Details for explicitly calculating the constants $K_\Lambda$ will be discussed in \cref{SecAsym}. Of the twelve lattices considered here, the Triangular lattice, from which the Apollonian lattices are derived, has the smallest value of $K_\Lambda$, and therefore the highest critical temperature as a function of $\qmax$. This is also demonstrated in \cref{fig_tc_vs_qmax_with_bounds}, in which we plot $\Tc$ versus $\qmax$ for lattices from all twelve families. These values of $\Tc$ are described asymptotically by \cref{asympqmax}. As is apparent from the figure, for large $\qmax$, each family has $\Tc$-values which scale in the same way, differentiated only by a vertical shift by the constant $K_{\Lambda}$. We conjecture that no planar lattice $\Lambda$ in Euclidean space can have a smaller value of $K_\Lambda$ than that of the Triangular lattice $K_{\Delta}$.

\begin{table}[t!]
\centering
{\setlength{\tabcolsep}{3.5pt} 
\begin{tabular}{c|l|c||l|c}
& \multicolumn{2}{c||}{Apollonian Lattices}
& \multicolumn{2}{c}{Laves-CaVO family} \\
\hline
$n$ & Lattice $\Lambda_n$ & $\TcLn/J$ & Lattice $\Lambda_n$ & $\TcLn/J$ \\
\hline
\hline 
0 &  Triangular & 3.641 & Laves-CaVO & 3.931 \\
1 &  Laves-Star & 5.007 & Salt cellar & 5.327 \\
2 &  Compass-Rose & 6.492 & Diamond-Kite & 6.833 \\
3 & Spectacular & 8.062 & Ces\'aro Square & 8.419 \\
4 & Apollonian--$96$ & 9.697 & Laves-CaVO--$128$ & 10.07 \\
5 & Apollonian--$192$ & 11.38 & Laves-CaVO--$256$ & 11.76 \\
6 & Apollonian--$384$ & 13.10 & Laves-CaVO--$512$ & 13.49 \\
7 & Apollonian--$768$ & 14.85 & Laves-CaVO--$1024$ & 15.24 \\
8 & Apollonian--$1536$ & 16.63 & Laves-CaVO--$2048$ & 17.02 \\
9 & Apollonian--$3072$ & 18.42 & Laves-CaVO--$4096$ & 18.82 \\
10 & Apollonian--$6144$ & 20.24 & Laves-CaVO--$8192$ & 20.64 \\
\end{tabular}
}
\caption{Values of $\TcLn/J$ for the Apollonian lattices and for the Laves-CaVO family. The first four rows ($n=0,1,2,3$) show the critical temperatures of the lattices depicted in \cref{ItTrPanel}. Lattices constructed with $n\geq4$ iterations are denoted by Apollonian--$\qmax$ and Laves-CaVO--$\qmax$, respectively. For the Apollonian lattices, $\qmax=2^n6$, whereas for the Laves-CaVO family we have $\qmax=2^n8$. For large $n$, the computed critical temperatures are observed to grow linearly in $n$, as predicted by \cref{asympqmaxn}.}
    \label{tab_Tn}
\end{table}

\section{Asymptotics of iterative triangulation}\label{SecAsym}

In this section, we establish the asymptotic equality from \cref{asympqmax},
\begin{equation}
   \frac{\Tc^{\Lambda_n}}{J} = A \ln  \qmax^{\Lambda_n} - 2 \ln \ln  \qmax^{\Lambda_n} - K_\Lambda + o(1),
\end{equation}
which emerges under iterative triangulation.
Here $q_{\rm max}^{\Lambda_n}$ is the maximal coordination number of $\Lambda_n$ and $A=2/\ln2$. To derive the asymptotics, we first derive \cref{asympqmaxn} and then express our result in terms of $\qmax^{\Lambda_n}$. Denote by $x_n=\tc^{\Lambda_n}$. Applying $g$ to \cref{indtc}, we deduce the forward recursion
\begin{equation} \label{xnEq}
    x_{n+1}=g(x_n).
\end{equation}
We seek to evaluate the asymptotic nature of $x_n$ as $n\to \infty$. In order to do so, we first show that $x_n\to 0$ for large $n$. Then we expand $g$ in \cref{xnEq} about $x_n=0$ using its analytic expression from \cref{gdef}. Next, we use the so-called Stolz–Ces\`{a}ro (SC) \cite{Stolz1885,Cesaro1888} theorem to extract the leading asymptotics of $x_n$ through the difference $1/x_{n+1}-1/x_n$. We successively refine our asymptotics, extracting the leading $n$-dependent terms until the remainder is summable.

\begin{table}[t!]
    \centering
    \begin{tabular}{c|c||c|c}
        {Base Lattice $\Lambda$} & {$\qmax^\Lambda$} & {\quad$\Tc^\Lambda/J$\quad} & {\quad$K_{\Lambda}$\quad} \\
        \hline
        \hline
        Triangular & 6 & 3.641 & 1.024 \\
        Half-Interpolated Triangular & 9 & 4.404 & 1.051 \\
        Spotted Triangular & 7 & 3.845 & 1.156 \\
        SrCuBO--7 & 7 & 3.810 & 1.209 \\
        Tri-Hexagonal & 8 & 4.051 & 1.231 \\
        Striped Triangular & 8 & 4.040 & 1.247 \\
        SrCuBO--10 & 10 & 4.445 & 1.296 \\
        SrCuBO--8 & 8 & 3.976 & 1.343  \\
        Laves-CaVO & 8 & 3.931 & 1.411  \\
        SrCuBO--9 & 9 & 4.128 & 1.457  \\
        Split Brick & 7 & 3.282 & 2.035  \\
        Laves-SHD & 12 & 4.136 & 2.274 
    \end{tabular}
    \caption{ 
    Table listing the critical temperatures $\TcL$ for all twelve base lattices $\Lambda$ considered in this work, obtained from the Kac--Ward formula in \cref{KacWardTc}. Under iterative triangulation, the critical temperatures for a given family of lattices $\Lambda_n$ grow asymptotically with the maximal coordination number $\qmax$ according to the universal scaling law from \cref{asympqmax}, with an additional negative offset by the lattice-dependent constant $K_\Lambda$. Lattices in this table are sorted by the value of $K_\Lambda$, where smaller values correspond to larger critical temperatures as a function of $\qmax$. The Triangular lattice, from which the Apollonian lattices are derived, has the smallest value of $K_{\Lambda}$ among the lattices considered here, indicating that the Apollonian lattices have the fastest-growing critical temperature with $\qmax$.}
    \label{tab_base_lat_data}
\end{table}

To show that $x_n\to0$, we use the fact that $g$ is bounded below by $0$ and that for fixed $x\in(0,1$), the sequence $g^n(x)$ is strictly decreasing. Since $g$ contains exactly two fixed points $x=0$ and $x=1$, we conclude that $x_n\to0$ as $n\to \infty$. For a proof of the monotonicity of $g^n(x)$ and the fact that $x_n\to 0$, we refer to \cref{monotoneg}. Now we use the SC theorem, which states that if $(a_n)_{n\in\mathbb{N}}$ is a sequence of real numbers and $(b_n)_{n\in\mathbb{N}}$ is a strictly monotone divergent sequence such that 
\begin{equation}
    \lim_{n\to \infty}\frac{a_{n+1}-a_{n}}{b_{n+1}-b_{n}}=L
\end{equation}
exists, then 
\begin{equation}
    \lim_{n\to \infty}\frac{a_n}{b_n} = L
\end{equation}
for $L\in\mathbb{R}$ \cite{Muresan2008}.
Using the SC theorem on the sequences $a_n= 1/x_n$ and $b_n= n$, we derive 
\begin{align}\label{SC1}
    \nonumber \lim_{n\to \infty} \frac{\frac{1}{x_{n}}}{n}&\stackrel{\rm SC}{=} \lim_{n\to \infty} \frac{\frac{1}{x_{n+1}}-\frac{1}{x_{n}}}{(n+1)-n}\\&=\lim_{n\to \infty}\left(\frac{1}{x_{n+1}}-\frac{1}{x_{n}}\right).
\end{align}
To evaluate the resultant difference, we expand $g$ about the fixed point $x=0$ and find
\begin{equation}\label{gexpEq}
    x_{n+1}=g(x_n)=x_n- 2 x_n^2+8x_n^3-36x_n^4 +\mathcal{O}(x_n^5).
\end{equation}
\Cref{gexpEq} can then be used to obtain the difference
\begin{equation}
    \frac{1}{x_{n+1}}-\frac{1}{x_{n}}=2-4x_n+12 x_n^2+\mathcal{O}(x_n^3),
\end{equation}
whose limit gives
\begin{equation}
    \lim_{n\to \infty}\frac{\frac{1}{x_n}}{n}=2.
\end{equation}
This implies that 
\begin{equation}\label{1/x1}
    \frac{1}{x_n}=2n+o(n),
\end{equation} 
where $o(n)$ denotes little-o. For arbitrary functions $f$ and $g$, ${f\in o(g(n))}$ implies that ${\lim_{n\to \infty}f(n)/g(n)=0}$ \cite{Muresan2008}. Here, by ${2n+o(n)}$, we mean ${2n+f(n)}$ for some $f\in o(n)$.
Inverting \cref{1/x1} gives
\begin{equation}\label{x1}
    x_n=\frac{1}{2n}+o\left(\frac{1}{n}\right).
\end{equation}

In order to extract the next-to-leading term in the asymptotic expansion of $x_n$, we use our result from \cref{x1}, together with SC for the sequences ${a_n=1/x_n-2n}$ and ${b_n=-2\ln n}$, to obtain 
\begin{align}
    \nonumber \lim_{n\to \infty}\frac{\frac{1}{x_n}-2n}{-2\ln n}&\stackrel{\rm SC}{=}\lim_{n\to \infty}-\frac{1}{2}\frac{\frac{1}{x_{n+1}}-\frac{1}{x_{n}}-2}{\ln(n+1)-\ln n}\\\nonumber
    &=\lim_{n\to \infty}\frac{\frac{1}{n}}{\ln(1+\frac{1}{n})}
+ o\left(\frac{1}{n \ln\left(1+\frac{1}{n}\right)}\right)\\&=1, \label{next-to-leading} \end{align} 
where we have substituted the result for $x_n$ from \cref{x1} into
\begin{equation}\label{diff}
    \frac{1}{x_{n+1}}-\frac{1}{x_{n}}-2= -4 x_n+12x_n^2 +\mathcal{O}(x_n^3).
\end{equation}
We used the fact that the sum over $n$ of the terms $1/x_{n+1}-1/x_n-2$, which grows as $1/x_n-2n$ for $n\gg 1$, simultaneously grows logarithmically according to \cref{diff} (since $\sum_{n}-4x_n\sim -2\sum_n 1/n$) in order to set the term in the denominator to $-2\ln n$. 
\cref{next-to-leading} implies that
\begin{equation}\label{firstitr}
    \frac{1}{x_n}=2(n-\ln n)+ o(\ln n).
\end{equation}
Inverting for $x_n$, we find that it is given by 
\begin{equation}\label{x2}
    x_n=\frac{1}{2(n-\ln n)}+ o\left(\frac{\ln n}{(n-\ln n)^2}\right).
\end{equation}
At last, substituting \cref{x2} into \cref{diff} yields
\begin{equation}\label{xn+1-xn}
    \frac{1}{x_{n+1}}-\frac{1}{x_{n}}= 2\left(1-\frac{1}{n}\right)+ \frac{3-2\ln n}{n^2}+\mathcal{O}\left(\frac{1}{n^3}\right).
\end{equation}
The difference $1/x_{n+1}-1/x_{n}-2(1-1/n)$ is now summable since the infinite series $\sum_{n\geq 1} \ln n/n^2$ and $\sum_{n\geq 1} 1/n^p$ for $p\geq2$ are finite. This implies that the next term in \cref{firstitr} is a constant with $o(1)$ corrections
\begin{equation}
 \frac{1}{x_{n}}= 2(n-\ln n)+\kappa_\Lambda+ o(1),  
\end{equation}
for some constant $\kappa_\Lambda>0$. However, since 
\begin{equation}
    \frac{1}{x_n}= \frac{1}{g^n(\tc^\Lambda)},
\end{equation}
we infer that $\kappa_\Lambda$ depends on the initial condition ${x_0=\tc^\Lambda}$. It is given by $\kappa_\Lambda=\kappa(\tc^\Lambda)$, where $\kappa(t)$ is defined as
    \begin{equation}\label{kappa}
    \kappa(t) :=\lim_{n\to \infty}\left(\frac{1}{g^n\left(t\right)}-2(n-\ln n)\right).
\end{equation}
The proof of the existence of the limit in the definition of $\kappa(t)$  for any $t\in(0,1)$ is given in \cref{App:kappa}, alongside numerical values of $\kappa(t_{\rm c}^\Lambda)$ for the base lattices in \cref{fig_base_lattices}.
Since $x_n=\tc^{\Lambda_n}$, we have that the critical temperature variable satisfies
\begin{equation}\label{1/tc_of_n}
    \frac{1}{\tcLn}= 2(n-\ln n )+\kappa(\tcL) + o(1),
\end{equation}
or
\begin{equation}\label{tc_of_n}
    \tc^{\Lambda_n}= \frac{1}{2(n-\ln n)+\kappa(\tc^\Lambda)} + o\left(\frac{1}{n}\right),
\end{equation}
which implies through \cref{tctanh} that
\begin{equation}
    \frac{\Tc^{\Lambda_n}}{J} = 2( n - \ln n ) + \kappa(\tc^\Lambda) + o(1)\,, \label{Tc_asymptotic}
\end{equation}
proving \cref{asympqmaxn}. This asymptotic behavior for large $n$ can be seen explicitly in \cref{tab_Tn}.

In order to rewrite this in terms of the maximal coordination number and show \cref{asympqmax}, we  recall from \cref{qmaxn} that $\qmax^{\Lambda_n} = 2^n\qmax^{\Lambda}$, which can be used to deduce
\begin{equation}\label{n_vs_q}
     n = \frac{1}{\ln 2}\ln \frac{\qmax^{\Lambda_n}}{\qmax^{\Lambda}} \,.
\end{equation}
Using this allows us to express the logarithmic term in \cref{Tc_asymptotic} as
\begin{align}\label{lnntoq}
     \ln n&=-\ln\ln 2+ \ln \ln \qmaxLn+o(1),
\end{align}
where we have used the fact that for large $\qmaxLn$,
\begin{equation}\label{lnlno1}
    \ln \ln \left(1-\frac{\qmaxLz}{\qmaxLn}\right)=o(1).
\end{equation}
Combining these results, \cref{Tc_asymptotic} becomes
\begin{equation}\label{Tc_asymp2}
\begin{aligned}
    \frac{\Tc^{\Lambda_n}}{J} = & A \ln \qmax^{\Lambda_n} - 2\ln \ln \qmax^{\Lambda_n} - K_{\Lambda} + o(1),
\end{aligned}
\end{equation}
where $A=2/\ln2$ and where we define the base-lattice-dependent constant
\begin{equation}
    K_\Lambda := A \ln  (\qmax^{\Lambda}) - \kappa(\tc^{\Lambda}) - 2\ln\ln2\label{K_Lamdbda1}
\end{equation}
using $\kappa$ as defined in \cref{kappa}
The plot in the right panel of \cref{fig_tc_vs_qmax_with_bounds} shows how each base lattice gives rise to a different value of $K_\Lambda$, which has the effect of an overall shift in the critical temperatures of the lattices in the corresponding family.
Similarly, we may write $\tcLn$ in terms of $\qmaxLn$ as
\begin{equation}\label{tclnasymp}
    \tcLn= \frac{1}{A \ln \qmaxLn-2 \ln \ln \qmaxLn-K_{\Lambda}}+ o\left(\frac{1}{\ln \qmaxLn}\right).
\end{equation}

\section{Continuous extension of $\TcLn$ and $\Tc^*(q_{\rm max})$}\label{comparison}

In this section, we construct the unique continuous extension for $T_{\rm c}^\Lambda(q_{\rm max})$, which satisfies \cref{Tc_asymptotic} and smoothly interpolates the $T_{\rm c}$-values for all members of the family derived from a given base lattice $\Lambda$. Furthermore, we construct it such that it asymptotically behaves as Eq. (\ref{Tc_asymp2}), namely
\begin{equation}
    \frac{\Tc^\Lambda(\qmax)}{J} =  A \ln \qmax - 2\ln \ln \qmax - K_{\Lambda} + o(1).
\end{equation}
We motivate our construction in the spirit of comparing the critical temperatures between various families at fixed $\qmax$. Since different base lattices may have different starting $\qmaxLz$, their corresponding families will have incommensurate ${\qmaxLn=2^n \qmaxLz}$, obstructing a direct comparison of their critical temperatures. The extension formula, however, circumvents this problem as it can be applied for all $q_{\rm max}>0$. We then construct the function 
\begin{align}
 T_{\rm c}^*(q_{\rm max}) = T_{\rm c}^\Delta(q_{\rm max}),
\end{align}
where the base lattice is chosen as the Triangular lattice, $\Lambda=\Delta$. We find that the curve $T_{\rm c}^*(q_{\rm max})$, which interpolates the $T_{\rm c}$-values of the Apollonian lattices, bounds all other lattices in this work, see \cref{fig_tc_vs_qmax_with_bounds}.

We construct the function $\Tc^\Lambda(\qmax)$ for a particular family in terms of an equivalent critical weight $\tau_{\rm c}^\Lambda(\qmax)$ which satisfies
\begin{equation}
    \frac{\Tc^\Lambda(\qmax)}{J}=\frac{1}{\arctanh\bigl(\tau_{\rm c}^{\Lambda}(\qmax)\bigr)}.
\end{equation}
Thus, such a continuous extension of $\TcLn$ exists if there exists a unique continuous extension $\tau_{\rm c}^{\Lambda}(q)$ for the critical weights $\tcLn$. Such a continuous extension must simultaneously satisfy the asymptotic relation \cref{tclnasymp} and the defining property \cref{indhtc}, which implies that it is only defined by the properties
\begin{align}\label{taup1}
    \tau_{\rm c}^\Lambda(q)&=\frac{1}{A \ln q-2\ln\ln q-K_{\Lambda}}+o\left(\frac{1}{\ln q}\right),\\ \label{taup2}
    \tau_{\rm c}^\Lambda(q)&=h^n\bigl(\tau_{\rm c}^\Lambda(2^n q)\bigr) ,
\end{align}
for $q>0$ and $n\in\mathbb{N}_0$. Indeed, we can explicitly construct such a continuous extension via
\begin{equation}\label{taucK}
    \tau_{\rm c}^{\Lambda}(q)= \lim_{n\to \infty}h^n\Bigl(\Bigl[A\ln \bigl(2^nq \bigr) - 2\ln \ln \bigl(2^nq \bigr) - K_\Lambda\Bigr]^{-1}\Bigr),
\end{equation}
where $K_{\Lambda}$ is the associated constant for the base lattice $\Lambda$ defined in \cref{K_Lamdbda1}. A proof of the existence and uniqueness of $\tau_{\rm c}^\Lambda$ for any base lattice $\Lambda$ is given in \cref{App:tauK}. 

In particular, when $\Lambda=\Delta$, we denote the continuous extension of $\Tc^\Delta(q)$ for the Apollonian lattices as $\Tc^*(q)$ defined through
\begin{equation}\label{Tstarsec}
    \frac{\Tc^*(q)}{J}=\frac{1}{\arctanh\bigl(\tc^*(q)\bigr)},
\end{equation}
where $\tc^*(q)= \tau_{\rm c}^\Delta(q)$ is given by 
\begin{equation}\label{tstarsec}
    \tc^*(q) = \lim_{n\to\infty} h^n \Bigl( \Bigl[A\ln \bigl(2^nq \bigr) - 2\ln \ln \bigl(2^nq \bigr) - K_\Delta \Bigr]^{-1} \Bigr),
\end{equation}
using $K_{\Delta}=1.024$. Numerical values of $\Tc^*$ for various $\qmax$ are given in \cref{tab_tstar}. In this table, we also show which of the lattices considered in this work, if any, have the largest $\Tc$ for a given $\qmax$. Together, \cref{fig_tc_vs_qmax_with_bounds} and \cref{tab_tstar}
highlight the fact that $\Tc^*(q)$ serves as an upper bound for the critical temperatures of all base lattices considered in this work.

\begin{table}[tb]
    \centering
    \begin{tabular}{c||c|c|c}
         {$\qmax$} & {$\Tc^*/J$} & {Maximal $\Tc/J$} & Maximal Lattice \\
         \hline 
         \hline
         6  & 3.641 & 3.641 & Triangular \\
         7  & 3.932 & 3.845 & Spotted Triangular \\
         8  & 4.191 & 4.051 & Tri-Hexagonal \\
         9  & 4.423 & 4.404 & Half-Interpolated Triangular \\         
         10 & 4.635 & 4.445 & SrCuBO--10 \\
         11 & 4.828 & 4.452 & Rep-11 \\
         12 & 5.007 & 5.007 & Laves-Star \\
         13 & 5.173 & {---} & {---} \\
         14 & 5.328 & 5.232 & Spotted Triangular--14 \\          
         15 & 5.474 & {---} & {---} \\         
    \end{tabular}
    \caption{Values of $\Tc^*$ for $\qmax$ between $6$ and $15$. The third column shows the largest critical temperature found among those lattices considered in this work with the given value of $\qmax$. The Rep-11 lattice \cite{Joseph2026} is not a triangulation, but is included here since we do not consider any lattices with $\qmax=11$. The lattice listed for $\qmax=14$ is obtained from the Spotted Triangular lattice via iterative triangulation. While the Triangular lattice with $\qmax=6$ and the Laves-Star lattice with $\qmax=12$ saturate $\Tc=\Tc^*$, all other lattices considered in this work have critical temperatures which fall short of $\Tc^*$. We conjecture that no planar tessellation with maximal coordination number $\qmax$ can possess a critical temperature $\Tc>\Tc^*(\qmax)$ in Euclidean space.}
    \label{tab_tstar}
\end{table}
In practice, computing $\Tc^*(q)$ using \cref{Tstarsec,tstarsec} is numerically  inefficient since the $o(1/\ln q)$ corrections in \cref{taup1} decay very slowly. In \cref{app_Tc_star}, we present an alternate representation of $\Tc^*(q)$ as a limit which converges rapidly in $n$. We also present the first few coefficients for a Taylor expansion of $\Tc^*$ about $\qmax=q=15$. The expansion gives accurate numerical values of $\Tc^*$ for $6\leq \qmax\leq 29$ with a relative error less than one percent.

\section{Conclusion and Outlook}

In this work, we have studied the ferromagnetic Ising model on two-dimensional lattices with the explicit goal of finding lattices with large critical temperatures $\Tc$. We devised a systematic method for constructing high-$\Tc$ lattices through the process of iterative triangulation, which can be used to generate lattices with arbitrarily high $\Tc$. We find that iterative triangulation gives critical temperatures that grow as $\Tc/J\sim\ln\qmax$, where $\qmax\gg1$ is the maximal coordination number of the lattice, in contrast to the linear scaling exhibited by the exact upper bound $\Tc\lesssim(2/\pi)\qmax$  previously computed in Ref.~\cite{Joseph2026}. As an additional benefit, our procedure allows us to compute analytic expressions for thermodynamic variables for entire families of lattices if the corresponding quantity for the base lattice is known.

In particular, iterative triangulation can be used on the Triangular lattice to generate the Apollonian lattices, a family of lattices whose critical temperatures are the highest, as a function of $\qmax$, of all lattices studied in this work. We conjecture that the Apollonian lattices are optimal among planar lattices in Euclidean space in the sense of achieving the largest possible critical temperature for a given $\qmax$. We denote this possibly maximal temperature by $\Tc^*(\qmax)$ and have presented an explicit expression for $\Tc^*$ for all $\qmax$. This represents a concrete criterion for testing the conjecture that the Apollonian lattices are optimal: to find a counterexample one must construct a planar lattice with some $\qmax$, whose critical temperature $\Tc$ lies above our conjecture. We have explicitly constructed twelve families of high-$\Tc$ lattices and computed their critical temperatures. None of the lattices under consideration have critical temperatures exceeding $\Tc^*$.

Future directions include extending our work to include periodic but non-isotropic ferromagnetic coupling $J_{ij}$. Under iterative triangulation, one can examine the behavior of critical temperatures for coupling strengths $J_{ij}$ which depend on the distance between neighbouring sites $i$ and $j$ connected by an edge on the lattice, or for strengths $J_n$ which depend only on the iteration $n$ at which a bond was introduced. Such variable coupling schemes might serve as a suitable framework for experimentally realizing Coherent Ising Machines for different topologies on the plane, since the modular nature of the lattice (in contrast to all-to-all coupling) is more manageable from an engineering perspective. In addition, the iterative triangulation procedure outlined in this work can be extended to include site-dependent coupling strengths provided the periodicity of the underlying lattice is preserved. Additionally, investigating the antiferromagnetic Ising model on these lattices lends itself to exact expressions for thermodynamic quantities like the free energy per site or residual entropy using our method and can serve as a framework for investigating frustration on lattices from these families. Moving away from the Ising constraint $s_i^2=1$, the dynamics of both non-interacting and interacting real scalar fields $\phi_i$ on artificial lattices has been demonstrated experimentally in topoelectrical circuits \cite{PhysRevLett.114.173902,PhysRevX.5.021031,lee2018topolectrical,kotwal2021active,Lenggenhager2021,zhang2022observation,chen2023hyperbolic,PhysRevResearch.5.L012041,Dey2024}. Therein, several hundreds of sites are easily achievable for planar lattices such as the ones proposed in this work.

The plethora of triangulations studied in this work, some of which have moderately small $\qmax$, can be used to construct new sets of lattices consisting of their duals. 
Since a triangulation consists only of triangular faces, the dual to any triangulation is made entirely of three-coordinated sites.
The presence of three-coordinated sites implies that these dual lattices might at first serve as a platform for the celebrated Kitaev quantum spin model. Originally  studied on the Honeycomb lattice, which is dual to the Triangular lattice \cite{Kitaev2006}, the Kitaev model has also been studied on the 
Star or decorated Honeycomb lattice, which is dual to the Laves-Star lattice
\cite{Yao2007,Dusuel2008,dOrnellas2024}. Indeed, it is the case that the dual of any planar triangulation admits a Kitaev model because it is three-edge colorable. This is a consequence of using the four color theorem \cite{Appel1989} on planar triangulations in conjunction with Tait's theorem \cite{Tait1880}. Such a three-edge coloring is known as a Tait coloring, which ensures the validity of the Kitaev model on these lattices. Since the high-$T_{\rm c}$ lattices studied here have $t_{\rm c}\simeq 0$, their dual lattices with $t_{\rm c}^{(\rm dual)} = \frac{1-t_{\rm c}}{1+t_{\rm c}}\simeq 1$ have very small values of $T_{\rm c}/J$ \cite{Joseph2026}. This suppression of classical ordering might be beneficial for stabilizing quantum spin liquid ground states in quantum models. Recently, the Kitaev model has been studied on three-coordinated hyperbolic $\{p,3\}$ lattices \cite{Mosseri2025,PhysRevLett.134.256604,s25y-s4fj,mx1t-74dm}.

In the spirit of investigating non-Euclidean tessellations, we emphasize that the calculations performed in this work assume that lattices can be mapped to planar graphs but are otherwise completely general. As such, iterative triangulation can be applied to classes of hyperbolic lattices to achieve large values of $\Tc$ in the ferromagnetic Ising model. For instance, the well-studied $\{3,7\}$ lattice, which consists of seven triangles meeting at a vertex, is a triangulation with $\Tc/J=5.350$ \cite{Breuckmann2020}, well above the conjectured bound of $\Tc^*/J=3.932$ for a lattice with $\qmax=7$. This result is not a contradiction to our conjecture, which pertains only to Euclidean space. Our results suggest that the family of lattices arising from iterative triangulation on the $\{3,7\}$ lattice would have critical temperatures which lie above those of the Apollonian lattices. However, rather remarkably, the asymptotic scaling under iterative triangulation in this non-Euclidean space is identical to that which we derived in this work, because these lattices are planar. For the $\{3,7\}$ lattice, we have $K_\Lambda=-1.006<K_{\Delta}$. Extending studies of Ising models using iterative triangulation on hyperbolic $\{3,q\}$ lattices from Ref. \cite{Gendiar2012} can provide a platform for discovering an upper bound for $\Tc$ in hyperbolic space in the future.

Lastly, the introduction of $\Tc^*$ provides an explicit conjectured bound on the critical temperatures of planar lattices in Euclidean space. It is known that higher critical temperatures for $k$-uniform lattices with $k\leq 6$ are correlated with higher average coordination number $\bar{q}$ \cite{Portillo}. Triangulations in this sense are optimal because they saturate the bound $\bar{q}\leq 6$ for any planar lattice in Euclidean space. Of the many triangulations considered in this work, the Triangular lattice is the only $q=6$ regular tiling on the Euclidean plane, and it simultaneously saturates the exact bound in \cite{Joseph2026}. These observations lend support to our conjecture that the Apollonian lattices derived from the Triangular lattice are optimal. However, the mere existence of a single lattice with a critical temperature above $\Tc^*$ is sufficient to disprove the conjecture, thereby establishing a natural open question.

\begin{acknowledgments}
The authors thank Alexander Hickey, Jongjun M. Lee, Sourav Biswas, Tomáš Bzdušek, Frank Marsiglio, and  Michael Scherer for fruitful discussions. The authors acknowledge funding from the Natural Sciences and Engineering Research Council of Canada (NSERC) Discovery Grants RGPIN-2021-02534 and DGECR2021-00043 and from Quantum Horizons Alberta.
\end{acknowledgments}

\begin{appendix}
    
\section{Star-triangle identity}
\begin{figure}[tb]
    \centering
    \includegraphics[width=\linewidth]{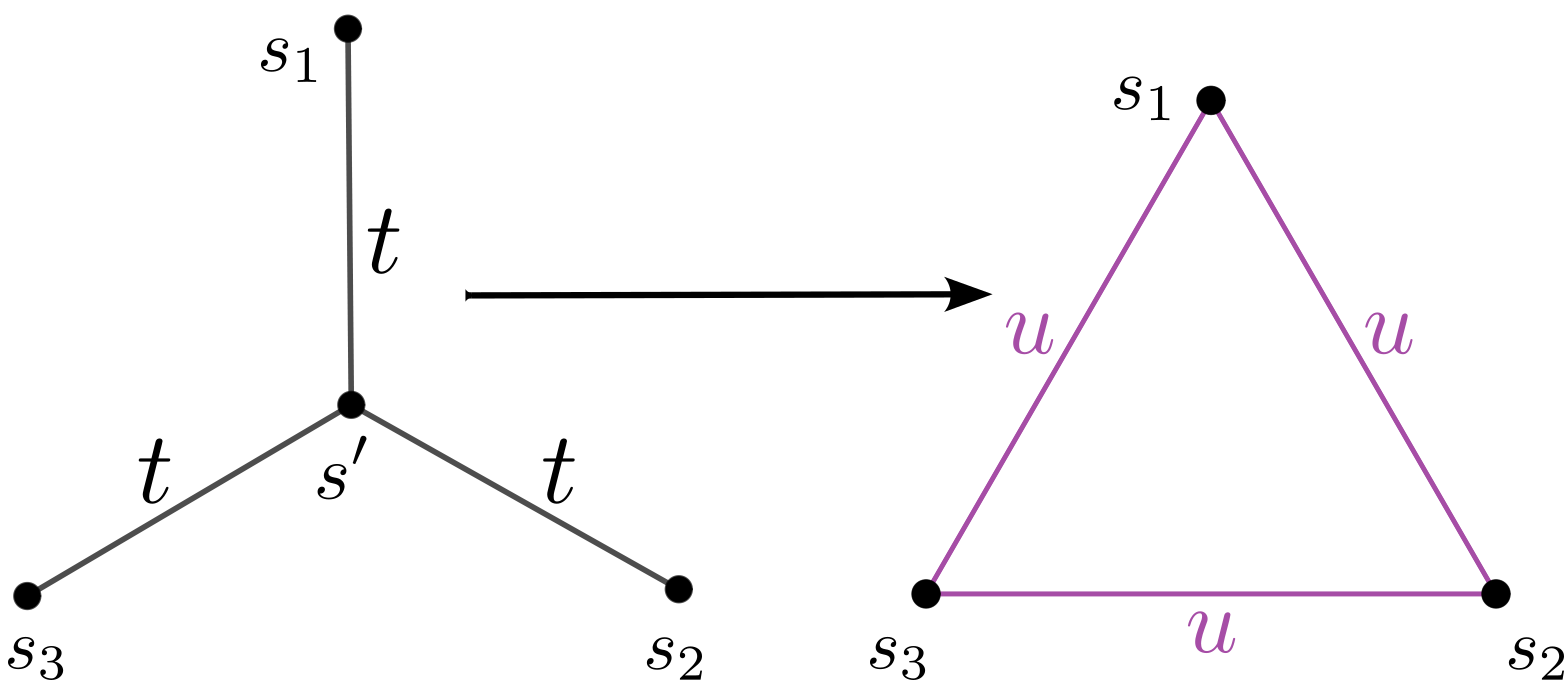}
    \caption{The star-triangle transformation allows for the decimation of a spin $s^\prime$ at the center of a star with coupling $t$ to spins $s_1,s_2,s_3$ that produces an effective  coupling $u$ (in purple) between the spins on a triangle.}
    \label{StTr}
\end{figure}

\label[appendix]{sttr}
In this section, we use the star-triangle identity on a generic star with a spin at its center to derive the functional form of the effective weight $u$. The star consists of spins $s_1,s_2,s_3$ on the legs alongside a spin $s^\prime$ located at the center connected to the other vertices as seen in \cref{StTr}. The edges between the spins carry a weight of $t=\tanh(\beta J)$ corresponding to the uniform coupling $J$. The partition function for the Ising model on this graph is of the form
\begin{equation}
    \mathcal{Z}\propto \sum_{s_1,s_2,s_3,s^\prime=\pm1}(1+t s_1 s^{\prime})(1+t s_2 s^{\prime}) (1+t s_3s^{\prime}).
\end{equation} 
The term within the sum expands to 
\begin{equation}\label{expandst}
\begin{aligned}
    &(1 + ts_1s') (1 + ts_2s') (1 + ts_3s')\\
    &\qquad\qquad = 1 + (ts_1s' + s_2s' + s_3s')\\ 
    &\qquad\qquad\qquad + t^2 \bigl( s_1(s')^2 s_2 + s_2 (s')^2s_3 + s_3 (s')^2 s_1 \bigr) \\ 
    &\qquad\qquad\qquad + t^3 s_1 s_2 s_3 (s')^3.
\end{aligned}
\end{equation}
Since $s^\prime=\pm1$, we infer $(s^\prime)^2=1$ and $(s^\prime)^3=s^\prime$. Together with the properties 
\begin{align}
\sum_{s^\prime=\pm 1} s^\prime&=0,\\
\sum_{s^\prime=\pm 1} 1&=2,
\end{align}
we decimate the spin $s^\prime$ from \cref{expandst} to deduce
\begin{multline}\label{decim_app}
    \sum_{s^{\prime}=\pm 1}(1+t s_1 s^{\prime})(1+t s_2 s^{\prime}) (1+t s_3s^{\prime})\\ =2 \bigl(1+t^2(s_1s_2+s_2s_3+s_3s_1) \bigr).
\end{multline}

Now we write \cref{decim_app} with the ansatz 
\begin{multline}
    2 \bigl(1+t^2(s_1s_2+s_2s_3+s_3s_1) \bigr) \\
    =C(1+u s_1s_2)(1+u s_2s_3)(1+u s_3s_1)
\end{multline}
in terms of an effective weight $u$ and a constant prefactor $C$, valid for all spin configurations of $s_1$, $s_2$, and $s_3$. Since the sum $s_1s_2+s_2s_3+s_3s_1$ evaluates to only two non-equivalent expressions for all configurations of $s_1$, $s_2$, and $s_3$, we have 
\begin{align}
    2(1+3t^2)&=C(1+u)^3,\label{cond1}\\
    2(1-t^2)&=C(1-u)^2(1+u).\label{cond2}
\end{align}
Using \cref{cond1,cond2}, we arrive at 
\begin{equation}
    \frac{1+3t^2}{1-t^2}=\frac{(1+u)^2}{(1-u)^2},
\end{equation}
which is solved for $u$ to produce
\begin{equation}\label{usol}
    u_{\pm}(t)=\frac{1+t^2\pm\sqrt{1+2 t^2-3 t^4}}{2 t^2},
\end{equation}
where we take the negative branch.
Using \cref{usol} in either \cref{cond1} or \cref{cond2} yields 
\begin{equation}
    C=\frac{16 t^6\left(3 t^2+1\right)}{\left(1+3 t^2-\sqrt{1+2t^2-3 t^4}\right)^3},
\end{equation}
where we identify ${C=G^{3/2}}$ from \cref{uGeq}. Setting $G=C^{2/3}$  produces \cref{Geq}.
\section{Monotonicity of $g^n(x)$ over $n$}\label[appendix]{monotoneg}

In this section, we prove that the functions $g^n(x)$ form a monotonically decreasing sequence in $n$ for fixed $x$ on the interval $(0,1)$. First, note that $g(x)$ is bounded below by $0$ by definition, i.e $g(x)>0$. The forward recursion in \cref{indtc} contains exactly two fixed points, $x=0$ and $x=1$, for the equation
    \begin{equation}
        x=g(x).
    \end{equation}
We will now show that $\forall x\in (0,1)$,
    \begin{equation}\label{gx_l_x}
        0<g(x)< x,
    \end{equation}
which implies that the next element in the sequence $x_{n+1}$ decreases as
\begin{equation}\label{gineq}
        x_{n+1}:=g(x_n)<x_{n}.
\end{equation}
To show \cref{gx_l_x}, we recall the definition of $g$ from \cref{gdef}, given by
\begin{equation}
    g(x)=\frac{2x}{1+x+\sqrt{1+6x-7x^2}},
\end{equation}
and observe that 
\begin{equation}
    1+6x-7x^2=(1+7x)(1-x).
\end{equation}
Since 
\begin{equation}
    1+7x > 1-x > 0
\end{equation}
for $0<x<1$, we have 
\begin{equation}
    1+x + \sqrt{1+6x-7x^2} > 1+x+\sqrt{(1-x)^2}=2.
\end{equation} 
Therefore, we have
\begin{equation}
    \frac{1}{1+x+\sqrt{1+6x-7x^2}}< \frac{1}{2},
\end{equation}
which gives the inequality
\begin{equation}
    g(x)=\frac{2x}{1+x+\sqrt{1+6x-7x^2}}<\frac{2x}{2}=x,
\end{equation}
demonstrating the claim.

\section{Induction proof for partition function and free energy per site}\label[appendix]{induction}

In this section, we provide proofs by induction for \cref{indz,indf,indhtc,indtc}.
Starting with \cref{indz}, the base case is established by \cref{forwardsZ}. Assume the induction hypothesis 
\begin{align}
    \nonumber \mathcal{Z}_{\Lambda_n}(t)=\mathcal{Z}_{\Lambda}\bigl(h^n(t)\bigr) \label{indzapp}\prod_{p=0}^{n-1}&\Biggl[\left(\frac{H\bigl(h^p(t)\bigr)G\bigl(h^p(t)\bigr)}{1- h^p(t)^2}\right)^{{V_{n}}/{3^p}} \\&\times \left(\frac{1-h^{p+1}(t)^2}{1-h^{p}(t)^2}\right)^{\frac{1}{2}{V_{n}}/{3^p}}\Biggr]
\end{align}
is true for $n\in\mathbb{N}$ and consider the lattice $\Lambda_{n+1}$ with edges $\mathcal{E}_{n+1}$ and with $V_{n+1}$ total vertices obtained via iterative triangulation on $\Lambda_{n}$. 
Performing the decimation over the new spins $s^\prime$ along the edges in $\mathcal{E}_{n+1}\backslash\mathcal{E}_{n}$ on $\Lambda_{n+1}$ using the same procedure that was used to derive \cref{forwardsZ}, we obtain
\begin{equation}\label{Z_nplus1}
    \mathcal{Z}_{\Lambda_{n+1}}(t)= \biggl(\frac{H(t) G(t)}{1-t^2}\biggr)^{V_{n+1}} \biggl(\frac{1-h(t)^2}{1-t^2}\biggr)^{\frac{1}{2}V_{n+1}}\mathcal{Z}_{\Lambda_n}(h(t)),
\end{equation}
which is simply \cref{forwardsZ}, but where we have set ${V_1\to V_{n+1}}$ and ${\mathcal{Z}_0\to\mathcal{Z}_{\Lambda_n}}$.
Taking the induction hypothesis \cref{indzapp} and replacing $t\to h(t)$, we deduce
\begin{align}\label{ZLn_ht}
    \nonumber \mathcal{Z}_{\Lambda_n}(h(t))=\mathcal{Z}_{\Lambda}(h^{n+1}(t)) \prod_{p=1}^{n}&\Biggl[\biggl(\frac{H(h^p(t))G(h^p(t))}{1- h^p(t)^2}\biggr)^{V_{n}/3^{p-1}} \\&\!\!\times \biggl(\frac{1-h^{p+1}(t)^2}{1-h^{p}(t)^2}\biggr)^{\frac{1}{2}V_{n}/3^{p-1}}\Biggr],
\end{align}
where we shifted the indices by $p\to p-1$. Recalling from \cref{3V} that $V_{n+1}=3V_{n}$, the exponent $V_{n}/3^{p-1}$ in \cref{ZLn_ht} can be re-expressed as $V_{n+1}/3^p$.
Combining this result with \cref{Z_nplus1} re-introduces the $p=0$ term, giving
\begin{align}
    \nonumber \mathcal{Z}_{\Lambda_{n+1}}(t)=\mathcal{Z}_{\Lambda}(h^{n+1}(t)) \prod_{p=0}^{n}&\Biggl[\biggl(\frac{H(h^p(t))G(h^p(t))}{1- h^p(t)^2}\biggr)^{V_{n+1}/3^p} \\&\!\!\times \biggl(\frac{1-h^{p+1}(t)^2}{1-h^{p}(t)^2}\biggr)^{\frac{1}{2}V_{n+1}/3^p}\Biggr],
\end{align}
thus proving \cref{indz}.

Taking the logarithm on both sides of \cref{indzapp} gives
\begin{align}\label{lnZLn}
     \nonumber  \ln\mathcal{Z}_{\Lambda_n}(t)&= \ln\mathcal{Z}_{\Lambda}\bigl(h^n(t)\bigr) \\& + \sum_{p=0}^{n-1} \frac{V_{n}}{3^p}\ln\left( \frac{H\bigl(h^p(t)\bigr)G\bigl(h^p(t)\bigr)}{1- h^p(t)^2}\right) \\ &+\frac{1}{2}\sum_{p=0}^{n-1} \frac{V_{n}}{3^p}\ln \left(\frac{1-h^{p+1}(t)^2}{1-h^p(t)^2}\right).
\end{align}
Dividing by $V_n$ and taking the thermodynamic limit $V_n\to\infty$ gives the free energy per site. We note that
\begin{equation}
\begin{aligned}
    \lim_{V_n\to\infty} \frac{\ln \mathcal{Z}_{\Lambda}\bigl(h^n(t)\bigr)}{V_n} &= \lim_{V_0\to\infty} \frac{V_0}{V_n}\frac{\ln \mathcal{Z}_{\Lambda}\bigl(h^n(t)\bigr)}{V_0} \\
    &= -\frac{1}{3^n} \beta f_\Lambda \bigl(h^n(t)\bigr),
\end{aligned}
\end{equation}
where we used \cref{Vn} alongside the definition of the free energy per site \cref{f_def}. Taken with \cref{lnZLn}, this gives
\begin{align}
    \nonumber -\beta f_{\Lambda_n}(t)=& 
    \lim_{V_n\to\infty} \frac{\ln \mathcal{Z}_{\Lambda_n}(t)}{V_n}=-\frac{1}{3^n}\beta f_{\Lambda}\bigl(h^{n}(t)\bigr)\\\nonumber &+\sum_{p=0}^{n-1}\frac{1}{3^p}\ln \left(\frac{H\bigl(h^p(t)\bigr)G\bigl(h^p(t)\bigr)}{1-h^p(t)^2}\right)\\ &+\frac{1}{2}\sum_{p=0}^{n-1}\frac{1}{3^p}\ln\left(\frac{1-h^{p+1}(t)^2}{1-h^p(t)^2}\right),
\end{align}
establishing \cref{indf}. Finally, given the free energy, the critical weight on $\Lambda_n$ now satisfies 
\begin{equation}
    h^n(\tc^{\Lambda_n})=\tc^\Lambda,
\end{equation}
which is inverted to show \cref{indtc}.

\section{Proof of existence of $\kappa(t)$ and numerical examples}
\label[appendix]{App:kappa}
In this section we prove the existence of $\kappa(t)$ as defined by the limiting procedure in \cref{kappa} for any $t\in(0,1)$ and provide values of $\kappa(\tc^\Lambda)$ for all the base lattices in \cref{fig_base_lattices}. To prove that the limit
\begin{equation}\label{kappa_def_app}
    \kappa(t)= \lim_{n\to \infty}\left( \frac{1}{g^n(t)} - 2(n-\ln n)\right)
\end{equation}
exists, let us first fix $t_0\in (0,1)$ and consider the sequence
\begin{equation}
    \kappa_n:= \frac{1}{\chi_n}-2(n-\ln n),
\end{equation}
where
\begin{align}
    \chi_n:=&g^n(t_0)\\
    \chi_{n+1}=&g(\chi_n)
\end{align}
for $n\in\mathbb{N}_0$. Observe that for $m\in \mathbb{N}_0$, the triangle inequality $|a+b|\leq |a|+|b|$ gives
\begin{equation}
    |\kappa_{m+1}-\kappa_1|\leq \sum_{n=1}^{m}|\kappa_{n+1}-\kappa_{n}|. 
\end{equation}
Observe that 
\begin{align}
    |\kappa_{n+1}-\kappa_n|
    &= \left|\frac{1}{\chi_{n+1}}-\frac{1}{\chi_{n}}-2+2\ln\left(1+\frac{1}{n}\right)\right|.
\end{align}
Since $\chi_n$ follows the exact same recurrence from \cref{xnEq}, it obeys \cref{xn+1-xn} which implies
\begin{equation}
    \frac{1}{\chi_{n+1}}-\frac{1}{\chi_{n}}=  2\left(1-\frac{1}{n}\right)+ \frac{3-2\ln n}{n^2}+\mathcal{O}\left(\frac{1}{n^3}\right).
\end{equation}
Therefore, 
\begin{align}
 \nonumber &|\kappa_{n+1}-\kappa_n|\\=& \left|-\frac{2}{n}+2 \ln\left(1+\frac{1}{n}\right)+\frac{3-2\ln n}{n^2}+\mathcal{O}\left(\frac{1}{n^3}\right)\right|.
\end{align}
For large $n$, the logarithm expands to
\begin{equation}
    2\ln\left(1+\frac{1}{n}\right)=\frac{2}{n}+\mathcal{O}\left(\frac{1}{n^2}\right),
\end{equation}
which implies that
\begin{equation}\label{kappan+1-n}
    |\kappa_{n+1}-\kappa_{n}|=\left|-2\frac{\ln n}{n^2}+\mathcal{O}\left(\frac{1}{n^2}\right)\right|\leq 2 \frac{\ln n}{n^2}+\mathcal{O}\left(\frac{1}{n^2}\right).
\end{equation}
Thus,
\begin{equation}
    \lim_{m\to \infty}|\kappa_{m+1}-\kappa_1| <\sum_{n=1}^{\infty}\mathcal{O}\left(\frac{\ln n}{n^2}\right)<\sum_{n=1}^{\infty}C\frac{\ln n}{n^2}
<\infty,
\end{equation}
for some constant $C>0$, proving that $\lim_{n\to \infty} \kappa_{n}$ exists. Since $t_0$ was arbitrary, we conclude that $\kappa(t)$ exists for all $t\in (0,1)$. Numerical values of $\kappa(t_{\rm c}^\Lambda)$ for all twelve base lattices shown in \cref{fig_base_lattices} can be found in \cref{kappavals}.

\begin{table}[tb]
    \centering
    \begin{tabular}{c|c|c|c}
        {Base Lattice $\Lambda$} & {$\Tc^\Lambda/J$} & {$\tc^\Lambda$} &  {$\kappa(\tc^\Lambda)$}  \\
        \hline
        \hline
        Triangular & 3.641 &  0.2679 & 4.879 \\
        Split Brick & 3.282 &  0.2956 & 4.313\\
        Laves-CaVO & 3.931 &  0.2490 &  5.322 \\
        Laves-SHD & 4.136 &  0.2372 &  5.629\\
        SrCuBO--7 &  3.810 & 0.2566 &  5.138 \\
        SrCuBO--8 &  3.977 & 0.2463 &  5.390\\
        SrCuBO--9 & 4.128 & 0.2380 &  5.616 \\
        SrCuBO--10 &   4.445 & 0.2212 & 6.081 \\
        Spotted Triangular &  3.845 &  0.2544 &  5.192  \\
        Tri-Hexagonal &  4.051 &  0.2420 & 5.502 \\
        Striped Triangular & 4.040 & 0.2426 &  5.486 \\
        Half-Interpolated Triangular &  4.404 & 0.2232 & 6.022  \\
    \end{tabular}
    \caption{Numerical values of $\kappa(t_{\rm c}^\Lambda)$ for the base lattices shown in \cref{fig_base_lattices}. Each base lattice $\Lambda$ is listed with its critical temperature $\Tc^\Lambda/J$ and critical temperature parameter $\tc^\Lambda$, the latter of which is used to compute $\kappa(\tc^\Lambda)$. Results converge to the given precision by truncating the limit in \cref{kappa_def_app} at $n=10^5$. }
    \label{kappavals}
\end{table}

\section{Existence and uniqueness of $\tauc{q}$}
\label[appendix]{App:tauK}
In this section, we prove that the unique continuous extension $\tauc{q}$ of $\tcLn$ for $q\geq 6$, which satisfies
\begin{align} \label{tauappp1}
\tau_{\rm c}^\Lambda(q)&=\frac{1}{A \ln q-2\ln\ln q-K_{\Lambda}}+o\left(\frac{1}{\ln q}\right),\\ \label{tauappp2}
\tau_{\rm c}^\Lambda(q)&=h^n(\tau_{\rm c}^\Lambda(2^n q)),
\end{align}
can be represented by the limit
\begin{equation}\label{taucKApp}
    \tauc{q}= \lim_{n\to \infty}h^n\Bigl(\Bigl[A\ln \bigl(2^nq \bigr) - 2\ln \ln \bigl(2^nq \bigr) - K_\Lambda \Bigr]^{-1}\Bigr),
\end{equation}
where 
\begin{align}
  K_{\Lambda}&= A \ln \qmaxLz- \kappa(\tcL)-2 \ln \ln 2
\end{align}
and $A=2/\ln2$. First, we prove that this limit exists and is finite for $q>0$. Define
\begin{equation}\label{F_C_q}
    F_C(q):= \lim_{n\to \infty}h^n\Bigl(\Bigl[A\ln \bigl(2^nq \bigr) - 2\ln \ln \bigl(2^nq \bigr) - C \Bigr]^{-1}\Bigr)
\end{equation}
for all $q>0 $ and fixed $C\in\mathbb{R}$. The existence of this limit implies that each curve $F_C(q)$ is uniquely parameterized by the constant $C$.  Let us define the sequence
\begin{equation}\label{bn}
    b_n:= \frac{1}{h^n\Bigl(\Bigl[A\ln \bigl(2^nq \bigr) - 2\ln \ln \bigl(2^nq \bigr) - C \Bigr]^{-1}\Bigr)}
\end{equation}
for $n\in\mathbb{N}$. Note that $b_n$ is well defined for sufficiently large $n$, since
\begin{equation}\label{bn>1}
    1\leq b_n<\infty
\end{equation}
when the quantity $A\ln \bigl(2^nq \bigr) - 2\ln \ln \bigl(2^nq \bigr) - C$ is positive. This is because $0<h(z)\leq1$ for all $z>0$, and consequently $0<h^p(z)\leq1$ for $p\in\mathbb{N}$. We will show that $\lim_{n\to \infty} b_{n}$ exists and is non-zero. Let $N\in\mathbb{N}_0$ and observe that the triangle inequality yields
\begin{equation}\label{bN}
    |b_{N+1}-b_{1}| \leq \sum_{n=1}^{N}|b_{n+1}-b_{n}|.
\end{equation}
In the following, we will show that $\exists n_0\in\mathbb{N}_0$ such that $\forall n> n_0$,
\begin{equation}\label{bn+1-bn}
    |b_{n+1}-b_n|\leq \mathcal{O}\left(\frac{\ln n}{n^2}\right),
\end{equation}
implying that 
\begin{equation}\label{sum_bn}
    \sum_{n=1}^{N}|b_{n+1}-b_n| <\sum_{n=1}^{\infty} \frac{\ln n}{n^2}<\infty.
\end{equation}
Performing the limit $N\to \infty$ on \cref{bN} gives
\begin{equation}\label{limbn}
    \lim_{N\to \infty }|b_{N+1}-b_1| < c,
\end{equation}
for some fixed $c>0$ showing that $\lim_{N\to \infty}b_{N}$ exists. 
To show \cref{bn+1-bn}, we require some necessary definitions. Fix $q_0> 0$ and define the sequence $\nu_{n}$ which satisfies
\begin{align}
    \label{nu1} \nu_n:=& \ln(2^n q_0),\\
    \label{nu2} \nu_{n+1}=& \ln(2)+ \nu_n,
\end{align}
and
\begin{equation}\label{zn}
    z_{n}:=A \nu_n-2\ln \nu_n - C.
\end{equation}
Using \cref{nu1,nu2,zn}, we express $b_n$ as 
\begin{equation}
    b_{n}= \frac{1}{h^n(z_n^{-1})}.
\end{equation}
Finally, define the auxiliary function $\eta(x)$ given by
\begin{equation}\label{aux1}
    \eta(z)= z-2+\frac{4}{1+z},
\end{equation}
and the reciprocal function
\begin{equation}
    r(z)=z^{-1},
\end{equation}
which satisfies
\begin{equation}\label{auxinv}
    r \circ z^{-1}=z.
\end{equation}
The auxiliary function $\eta$ is constructed to satisfy 
\begin{equation}\label{aux2}
    r\circ h= \eta\circ r.
\end{equation}
In particular for any $n\in\mathbb{N}_0$, applying \cref{aux2} repeatedly implies
\begin{equation}\label{aux3}
    b_n=r\circ h^{n} \circ z_{n}^{-1}= \eta^n \circ r\circ z_{n}^{-1}=\eta^n \circ z_{n},
\end{equation}
where we have employed the identity in \cref{auxinv}.
\Cref{aux3} allows us to express the difference $|b_{n+1}-b_{n}|$ as
\begin{align}
    \nonumber |b_{n+1}-b_{n}|&= \left|r\circ h^{n+1}\circ z_{n+1}^{-1}-r\circ h^n \circ z_{n}^{-1}\right|\\
&\nonumber =| \eta^{n+1} \circ z_{n+1}- \eta^n \circ z_{n}|\\
&=|\eta(\eta^{n}(z_{n+1}))-\eta(\eta^{n-1}(z_{n}))| .\label{bneta}\end{align}
We will now use the mean value theorem (MVT) on the function $\eta$ repeatedly to bound $|b_{n+1}-b_{n}|$ from above. The MVT implies that there exists $\xi\in (x,y)$ such that
\begin{equation}\label{MVT}
    |\eta(x)-\eta(y)| \leq |\eta^\prime(\xi)|\cdot |x-y|.
\end{equation}
The MVT applied once to $\eta$ guarantees the existence of $\xi\in (\xi^-,\xi^+):=(\eta^{n}(z_{n+1}),\eta^{n-1}(z_{n}))$ such that
\begin{equation}\label{MVT1}
    |\eta(\eta^{n}(z_{n+1}))-\eta(\eta^{n-1}(z_{n}))|\leq |\eta^\prime(\xi)| \cdot |\xi^- - \xi^+|.
\end{equation}
However, the value of $|\eta^{\prime}(\xi)|$ is unknown in general. We will show that if $\xi>1$, then 
\begin{equation}\label{eta'bound}
    |\eta^\prime(\xi)|<1.
\end{equation}
Observe that if $\xi>1$, then 
\begin{equation}
    |\eta^{\prime}(\xi)|= \left|1- \frac{4}{(1+\xi)^2}\right|<1.
\end{equation}
Furthermore, observe that for $z>1$
\begin{equation}
    \eta(z)-1=\frac{(1-z)^2}{1+z}>0,
\end{equation}
which implies that if $z>1$, then $\eta(z)>1$. The observation that $\eta(z)>1$ for $z>1$ readily generalizes for any $m\in\mathbb{N}_0$ into $\eta^m(z)>1$ for $z>1$. In particular, if we choose $n_0$ large enough such that $z_{n_0}>1$, then $\eta^{n-1}(z_n)=\xi^->1$ for all $n>n_0$ because $z_{n+1}>z_{n}$. Since $\xi^{-}<\xi<\xi^+$, we use \cref{eta'bound,MVT1} to deduce that \cref{bneta} is bounded above by
\begin{equation}
    |b_{n+1}-b_{n}|\leq |\eta(\eta^{n-1}(z_{n+1})-\eta(\eta^{n-2}(z_{n+1}))|.
\end{equation}
Applying the MVT $n-1$ times, we arrive at 
\begin{equation}\label{MVTex}
    |b_{n+1}-b_{n}| \leq | \eta(z_{n+1})-z_{n+1}|.
\end{equation}
Using the definition of the auxiliary function $\eta$, we have
\begin{align}
    \nonumber &|\eta(z_{n+1})-z_{n}|\\\nonumber &=\left|z_{n+1}-2 +\frac{4}{1+z_{n+1}}-z_{n}\right|\\
     &=\Biggl| 2 \ln\biggl(\frac{\nu_{n}}{\ln(2)+\nu_{n}}\biggr)+ \frac{4}{3+2\Bigl(\frac{\nu_n}{\ln(2)}-2 \ln(\ln(2)+\nu_n)\Bigr)}\Biggr|,
\end{align}
where we have used \cref{nu2}. Since $\lim_{n\to\infty} \nu_n\to \infty$, we expand about $1/\nu_n=0$ to obtain
\begin{align}
    \nonumber &2 \ln\left(\frac{\nu_{n}}{\ln(2)+\nu_{n}}\right)+ \frac{4}{3+2\left(\frac{\nu_n}{\ln(2)}-2 \ln(\ln(2)+\nu_n)\right)}\\
    \nonumber & =\frac{2}{\ln(2)^2}\left(\frac{2\ln(\nu_n)}{\nu_n^2}-\frac{1}{\nu_n^2}\right)+\mathcal{O}\left(\frac{1}{\nu_n^2}\right)\\
    \nonumber &=\mathcal{O}\left(\frac{\ln(\nu_n)}{\nu_n^2}\right).
    \end{align}
Since $\nu_n= n\left(\ln 2+\frac{\ln q_0}{n}\right)$, we deduce that 
\begin{align}
    &\frac{2}{\ln(2)^2}\left(\frac{2\ln(\nu_n)}{\nu_n^2}-\frac{1}{\nu_n^2}\right)+\mathcal{O}\left(\frac{1}{\nu_n^2}\right)\\&=\frac{4}{\ln(2)^4}\frac{\ln n}{n^2}+\mathcal{O}\left(\frac{1}{n^2}\right).
\end{align}
Therefore, we obtain 
\begin{equation}\label{eta_n+1-n}
    |\eta(z_{n+1})-z_{n}| = \frac{4}{\ln(2)^4}\frac{\ln n}{n^2}+\mathcal{O}\left(\frac{1}{n^2}\right).
\end{equation}
Thus, we conclude that $\forall n>n_0$
\begin{equation}\label{b_n+1-n}
    |b_{n+1}-b_{n}|\leq |\eta(z_{n+1})-z_{n}| = \frac{4}{\ln(2)^4}\frac{\ln n}{n^2}+\mathcal{O}\left(\frac{1}{n^2}\right),
\end{equation}
proving \cref{bn+1-bn} since $\ln n/n^2=\mathcal{O}(\ln n/n^2)$, and consequently implying that $\lim_{n\to \infty}b_{n}$ exists using \cref{sum_bn,limbn}. Moreover, $b_n$ exists for each $n$ since $z_n \neq 0,1$, so we can invert $b_n$ to conclude that $F_{C}(q_0)$ exists. Since $q_0> 0$ was arbitrary, we conclude that $F_C(q)$ is a well-defined function on $q\in (0,\infty)$.

Now we prove that $\tauc{q}$ is the only unique curve defined over $q\geq6$ which satisfies \cref{tauappp1,tauappp2} and is expressed as the limit
\begin{equation}\label{taucL}
\tauc{q}=F_C(q),
\end{equation}
where $C=K_{\Lambda}$. First we show that $F_C(q)$ satisfies the following key identity
\begin{equation}
    h(F_C(2q))=F_C(q).
\end{equation}
To show this property, we define the functions
\begin{equation}
    \zeta_{n}(q):=A \ln(2^n q)-2\ln \ln(2^nq) - C
\end{equation}
and
\begin{equation}
 \alpha_n(q)=h^n(1/\zeta_n(q))
\end{equation} 
for $n\in\mathbb{N}_0$. Observe that 
\begin{align}
    \nonumber h(\alpha_n(2q))&= h^{n+1}\left(\frac{1}{\zeta_n(2q)}\right)\\
    &=h^{n+1}\left(\frac{1}{\zeta_{n+1}(q)}\right).
\end{align}
On performing the limit $n\to \infty$ and using the continuity of $h$, we obtain
\begin{align}
    \nonumber \lim_{n\to \infty}h(\alpha_n(2q))&= \lim_{n\to \infty}h^{n+1}\left(\frac{1}{\zeta_{n+1}(q)}\right),\\
    \nonumber h\left(\lim_{n\to\infty}\alpha_n(2q)\right)&=\tauc{q},\\
    h(F_C(2q))&=F_C(q).
\end{align}
Using induction, we deduce that
\begin{equation}\label{hFK}
    h^m(F_C({2^m q}))= F_C(q)
\end{equation}
for $m\in \mathbb{N}_0$ and for $C=K_{\Lambda}$.

Now to prove \cref{taucL}, fix $q_0\geq 6$. For $n\in\mathbb{N}_0$ define the sequences 
\begin{align}
    \tau_n&:=\tauc{2^n q},\\
    f_n&:=F_{C}(2^n q).
\end{align}
In this notation, \cref{taucL} is equivalent to  
\begin{equation}
    \tau_0=f_0.
\end{equation}
Using \cref{hFK}, we deduce that 
\begin{align}
    h(f_{n+1})&=f_n,
\end{align}
and, from \cref{tauappp2}, we recall
\begin{equation}
    h(\tau_{n+1})=\tau_n,
\end{equation}
which together imply $\lim_{n\to \infty} \tau_n=0$ and $\lim_{n\to \infty} f_n=0$, following \cref{monotoneg}. We now consider the difference
\begin{equation}
    d_n= \tau_n^{-1}- f_n^{-1}.
\end{equation}
From \cref{tauappp2,hFK}, we can rewrite $\tau_n$ and $f_n$ as
\begin{align}
    \tau_n&=h^{m-n}(\tau_m),\\
    f_n&=h^{m-n}(f_m)
\end{align}
for some $\mathbb{N}_0\ni m>n$. Using these relations, we have
\begin{align}
    |d_n|&= \left|\frac{1}{h^{m-n}(\tau_m)}-\frac{1}{h^{m-n}(f_m)}\right|\\
    &=\left|\eta^{m-n} \circ \tau_m^{-1}- \eta^{m-n}\circ f_m^{-1}\right|,
\end{align}
where we have used the auxiliary function $\eta$ defined in \cref{aux1}. Recall that since both $\tau_m$ and $f_m$ approach $0$ as $m\to \infty$, $\exists m_0\in \mathbb{N}_0$ such that that both $\tau_m^{-1}$ and $f_m^{-1}$ are both greater than unity when $m>m_0$. Thus, we may employ the mean value theorem (MVT) $m-n$ times as identically done in \cref{MVTex} to obtain 
\begin{equation}
    |d_n|\leq |\tau_m^{-1}- f_m^{-1}|,
\end{equation}
where we used $|\eta^\prime|<1$ when $\tau_m^{-1},f_m^{-1}>1$. Using the triangle inequality, we have 
\begin{equation}
    |d_n|\leq |\tau_m^{-1}-z_m|+ |z_{m}-f_m^{-1}|.
\end{equation}
From \cref{tauappp1}, we infer 
\begin{equation}
    \tau_n= z_n^{-1}+ o(n^{-1}),
\end{equation}
which implies that 
\begin{equation}
    |\tau_m^{-1}-z_m|=o(m^{-1}).
\end{equation}
Simultaneously, we observe that 
\begin{align}
    \nonumber f_m&= F_C(2^m q)\\ \nonumber &=\lim_{k\to \infty}h^k\Bigl(\Bigl[A\ln \bigl(2^{k+m}q \bigr) - 2\ln \ln \bigl(2^{k+m}q \bigr) - C \Bigr]^{-1}\Bigr)\\
    &= \lim_{k\to \infty}h^k(z_{k+m}^{-1}).
\end{align}
This observation allows us to write $f_m^{-1}$ in terms of $\eta$ through
\begin{equation}
    f_m^{-1}=\lim_{k\to \infty}\frac{1}{h^k(z_{k+m}^{-1})}=\lim_{k\to \infty} \eta^k(z_{k+m}).
\end{equation}
Together with the fact that $\lim_{k\to \infty} \eta^k(z_{k+m})$ admits the telescoping sum decomposition
\begin{align}
 \lim_{k\to \infty}\eta^k(z_{k+m})=&\sum_{p=1}^{\infty}(\eta^{p}(z_{p+m})-\eta^{p-1}(z_{p+m-1}))+z_{m},
\end{align}
we infer that
\begin{equation}
    |z_m-f_m^{-1}|= \Biggl|z_m-\Biggl(\sum_{p=1}^{\infty}\eta^{p}(z_{p+m})-\eta^{p-1}(z_{p+m-1})\Biggr)-z_m\Biggr|
\end{equation}
which implies
\begin{align}
    |d_n|&\leq o(m^{-1})+\left|\sum_{p=1}^{\infty}\eta^{p}(z_{p+m})-\eta^{p-1}(z_{p+m-1})\right|\\
    &\leq o(m^{-1})+\sum_{p=1}^{\infty}|\eta^{p}(z_{p+m})-\eta^{p-1}(z_{p+m-1})|
\end{align}
using the triangle inequality once more. Choose $m_0$ now large enough such that $z_{m+p}>z_{m_{0}}>1$ is also true. Then, we can employ MVT yet again on $\eta^{p}(z_{p+n})-\eta^{p-1}(z_{p+n-1})$ exactly $p-1$ times to obtain
\begin{align}
    |d_n|&\leq o(m^{-1})+\sum_{p=1}^\infty |\eta{(z_{p+m})-z_{p+m}|}\\
    &=o(m^{-1})+\sum_{p=m}^{\infty}|\eta(z_p)-z_p|,
\end{align}
where we have shifted the index so that the sum begins at $p=m$. Since we have already computed $|\eta(z_p)-z_p|$ in \cref{eta_n+1-n}, we use its expression to conclude that 
\begin{equation}
    |d_{n}|\leq o(m^{-1})+\sum_{p=m}^{\infty}\frac{\ln p}{p^2}.
\end{equation}
Since $m>m_0$ was arbitrary, we may choose arbitrarily large $m\gg1$ to obtain the vanishing tail 
\begin{equation}
    \lim_{m\to \infty} \sum_{p=m}^{\infty} \frac{\ln p}{p^2}=0
\end{equation}
due to the fact that $\sum_{p=1}^{\infty} \ln (p)/p^2$ converges. Ultimately, we have that 
\begin{equation}
    0\leq |d_n|\leq \lim_{m\to \infty} \left(o(m^{-1})+\sum_{p=m}^{\infty}\frac{\ln p}{p^2}\right)=0.
\end{equation}
We conclude that $d_n=0$ for arbitrary $n\in\mathbb{N}_0$ and $q_0\geq 6$, and in particular for $n=0$ we have
\begin{equation}
    \tau_0=f_0 \quad \blacksquare.
\end{equation}

\section{Alternate representations of $\Tc^*(q)$}\label[appendix]{app_Tc_star}

In this section, we present two alternative expressions for $t_{\rm c}^*(q)$ and hence
\begin{align}
 \frac{T_{\rm c}^*(q)}{J} = \frac{1}{\arctanh\bigl(t_{\rm c}^*(q)\bigr)}.
\end{align}
The first expression is a rapidly convergent expression that can be used to evaluate the exact value of $t_{\rm c}^*(q)$ for all $q$. The second expression is a Taylor polynomial approximation that is applicable to reasonable accuracy for the important regime of small $q\in[6,24]$.

\subsection{Rapidly convergent expression}

We first show that
\begin{multline}{\label{tc_star_numerical_1}}
    \tc^*(q) = \lim_{n\to\infty} h^n \Biggl( \Biggl[ \frac{1}{g^n(\tcL)} + \frac{2}{\ln2}\ln \frac{q}{6} \\
    - 2\ln \biggl( 1 + \frac{1}{n\ln2} \ln \frac{q}{6} \biggr) \Biggr]^{-1} \Biggr),
\end{multline}
which converges much faster in $n$ than the expression in \cref{tstarsec} and thus allows for efficient numerical computation of $\Tc^*$. First, we outline a proof of the equivalence of \cref{tstarsec,tc_star_numerical_1}, and then show why the representation in \cref{tc_star_numerical_1} converges more quickly. 

Beginning from \cref{1/tc_of_n}, we insert \cref{n_vs_q} to arrive at
\begin{equation}
\begin{aligned}
    \frac{1}{\tcLn} &= \frac{2}{\ln2}\ln \frac{\qmax^{\Lambda_n}}{\qmax^{\Lambda}} - 2\ln\biggl( \frac{1}{\ln2} \ln \frac{\qmax^{\Lambda_n}}{\qmax^{\Lambda}} \biggr) + \kappa_n,
\end{aligned}
\end{equation}
where we define the remainder
\begin{equation}
    \kappa_n := \frac{1}{\tcLn} - 2 \bigl( n - \ln n \bigr).
\end{equation} 
Consider the Apollonian lattices, for which $\qmaxLz=\qmax^\Delta = 6$. We recall that $\qmax^{\Delta_n}=2^n\qmax^\Delta$, and evaluate this expression instead at $2^nq$ for some $q\neq\qmax^\Delta$ to arrive at
\begin{equation}
\begin{aligned}
    \frac{1}{\tc^*(2^nq)} \approx& \frac{2}{\ln2}\ln \frac{2^nq}{\qmax^{\Lambda}} - 2\ln\biggl( \frac{1}{\ln2} \ln \frac{2^nq}{\qmax^{\Lambda}} \biggr) + \kappa_n \\
    =& 2n + \frac{2}{\ln2}\ln \frac{q}{\qmax^{\Lambda}} - 2\ln\biggl( n + \frac{1}{\ln2} \ln \frac{q}{\qmax^{\Lambda}} \biggr) + \kappa_n  \\
    =& 2(n-\ln n) + \frac{2}{\ln2}\ln \frac{q}{\qmax^{\Lambda}}+\kappa_n \\&-2\ln \biggl( 1 + \frac{1}{n\ln2} \ln \frac{q}{\qmax^{\Lambda}} \biggr).
\end{aligned}
\end{equation}
Here, we write $\approx$ because we have kept $\kappa_n$ fixed, assuming that for large enough $n$ the error introduced is small. From here, we re-introduce the definition of $\kappa_n$ and simplify to produce
\begin{equation}
\begin{aligned}
    \frac{1}{\tc^*(2^nq)} &\approx 2n + \frac{2}{\ln2}\ln \frac{q}{\qmax^{\Lambda}} - 2\ln n \\
    &\qquad-2\ln \biggl( 1 + \frac{1}{n\ln2} \ln \frac{q}{\qmax^{\Lambda}} \biggr) \\&\qquad+ \frac{1}{\tcLn} - 2n + 2\ln n \\
    &= \frac{1}{\tcLn} + \frac{2}{\ln2}\ln \frac{q}{\qmax^{\Lambda}} - 2\ln \biggl( 1 + \frac{1}{n\ln2} \ln \frac{q}{\qmax^{\Lambda}} \biggr).
\end{aligned}
\end{equation}
Recalling that $\tc^*(2^nq)=g^n\bigl(\tc^*(q)\bigr)$ and that $\tc^{\Delta_n}=g^n(\tc^\Delta)$, we thus have
\begin{multline}
    g^n\bigl(\tc^*(q)\bigr) \approx \Biggl[ \frac{1}{g^n(\tc^{\Delta})} + \frac{2}{\ln2}\ln \frac{q}{\qmax^{\Delta}} \\
    - 2\ln \biggl( 1 + \frac{1}{n\ln2} \ln \frac{q}{\qmax^{\Delta}} \biggr) \Biggr]^{-1}.
\end{multline}
Applying $h^n$ on both sides gives
\begin{multline}{\label{tc_star_numerical}}
    \tc^*(q) =\lim_{n\to\infty} h^n \Biggl( \Biggl[ \frac{1}{g^n(\tcL)} + \frac{2}{\ln2}\ln \frac{q}{\qmax^{\Delta}} \\
    - 2\ln \biggl( 1 + \frac{1}{n\ln2} \ln \frac{q}{\qmax^{\Delta}} \biggr) \Biggr]^{-1} \Biggr),
\end{multline}
where taking the limit restores equality. From here, we compute
\begin{equation}
    \frac{\Tc^*(q)}{J} = \frac{1}{\artanh\bigl(\tc^*(q)\bigr)}.
\end{equation}

To see why this representation is numerically beneficial, fix $q_0\geq 6$ and define 
\begin{equation}\label{un}
    u_{n}:=2(n-\ln n)+ A \ln \frac{q_0}{6}- 2\ln \left(1+\frac{1}{n \ln 2}\ln \frac{q_0}{6}\right),
\end{equation}
where $A=2/\ln 2$ and
\begin{equation}
    B_n= \frac{1}{h^n(u_n^{-1})}.
\end{equation}
Then we claim that for large $n$
\begin{equation}
    \left|B_{n+1}-B_n\right| \leq 4\frac{\ln n}{n^2}+ o\left(\frac{\ln n}{n^2}\right),
\end{equation}
in contrast to \cref{b_n+1-n} which has 
\begin{equation}
    |b_{n+1}-b_n| \leq \frac{4}{(\ln 2)^4} \frac{\ln n}{n^2} + o\left(\frac{\ln n}{n^2}\right),
\end{equation}
where $b_n$ is defined in \cref{bn}. Since $(\ln2)^{-4}>1$, we conclude that $B_n$ approaches its limit quicker than $b_n$ when $n\to \infty$.

The proof of the claim is identical to the proof of \cref{bn+1-bn} but instead of $z_n$ defined in \cref{zn}, we use $u_n$ defined in \cref{un}. Observe that on choosing $n$ large enough so that $u_n^{-1}>1$, we then have
\begin{align}
\left|B_{n+1}-B_{n}\right|&= \left|\eta^{n+1}\circ u_{n+1}- \eta^n \circ u_n\right|\\ &\leq \left|\eta(u_{n+1})-u_n\right|\\&=\left|u_{n+1}-2 + \frac{4}{1+u_{n+1}}-u_n\right|.    
\end{align}
Thus
\begin{align}
    \nonumber  &\left|u_{n+1}-2 + \frac{4}{1+u_{n+1}}-u_n\right|\\=& \nonumber  \Bigl|\kappa_{n+1}-\kappa_n+2 \ln \frac{n}{1+n}+ 2 \ln \left(1+\frac{\ln q_0/6}{n\ln 2}\right)\\& +\frac{2}{1+\Psi_n}-\ln \left(1+\frac{\ln q_0/6}{n \ln 2+\ln  2}\right)\Biggr|\\ 
    \nonumber \leq& |\kappa_{n+1}-\kappa_n| +\Biggl|2 \ln \frac{n}{1+n}+ 2 \ln \left(1+\frac{\ln q_0/6}{n\ln 2}\right)\\& +\frac{2}{1+\Psi_n}-\ln \left(1+\frac{\ln q_0/6}{n \ln 2+\ln  2}\right)\Biggr|
\end{align}
where 
\begin{align}
    \nonumber \Psi_n=&\kappa_{n+1}-2 \ln  \left(\frac{\ln  q_0/6}{n \ln  2+\ln 
   2}+1\right)\\&+2 (n+1-\ln  (n+1))+\frac{2 \ln  q_0/6}{\ln  2}.
\end{align}

\begin{table}[tb]
    \centering
    \begin{tabular}{c|c}
        $k$ & $a_k$ \\
        \hline
    0 & $5.474\phantom{}$ \\
    1 & $\phantom{.}0.1412$ \\
    2 & $-0.00421$ \\
    3 & $\phantom{-}0.00017$ \\
    4 & $-8\times10^{-6}$ \\
    5 & $\phantom{-}4\times10^{-7}$ \\
    6 & $-2\times10^{-8}$
    \end{tabular}
    \caption{Table of approximate Taylor coefficients for the series expansion of $\Tc^*(q)/J$ around $q_0=15$. Keeping terms up to fourth order generates results which are accurate to 1\% for $7\leq q\leq 27$. To sixth order, the expansion is accurate to $10^{-3}$ for $8\leq q\leq 24$.}
    \label{table_Taylor_coefficients}
\end{table}

From \cref{kappan+1-n}, we have
\begin{equation}
    |\kappa_{n+1}-\kappa_n| \leq 2 \frac{\ln n}{n^2}+\mathcal{O}\left(\frac{1}{n^2}\right).
\end{equation} 
Moreover, because $\kappa_{n+1}<\kappa_n$ and $\lim_{n\to\infty}\kappa_n=\kappa$, we have
\begin{equation}
    \kappa_{n+1}=\kappa+ \mathcal{O}\left(\frac{\ln (n+1)}{(n+1)^2}\right)\leq \kappa+C \frac{\ln (n+1)}{(n+1)^2}
\end{equation}
for some constant $C>0$. Combining these and expanding about $1/n=0$, we obtain
\begin{equation}
    |B_{n+1}-B_n|\leq 2\frac{\ln n}{n^2}+2\frac{\ln n}{n^2}+\mathcal{O}\left(\frac{1}{n^2}\right).
\end{equation}
Since $f\in \mathcal{O}(1/n^2)\implies f\in o(\ln n /n^2)$, we have 
\begin{equation}
    |B_{n+1}-B_n|\leq 4\frac{\ln n}{n^2}+o\left(\frac{\ln n}{n^2}\right).
\end{equation}
Since our results used arbitrary $q_0\geq 6$, we deduce that the representation in \cref{tc_star_numerical} is more efficient than \cref{tstarsec}.
In our numerics, we found that indeed \cref{tc_star_numerical} gives much faster numerical convergence than the expression in \cref{tstarsec}, though both are exact in the infinite limit. Practical implementation of \cref{tc_star_numerical} is therefore not too difficult, with results converging rapidly in $n$.

\subsection{Taylor Polynomial approximation}

We now present numerical values for the first six coefficients of the Taylor series expansion of $\Tc^*(q)$ about $q=15$. 
By expanding about $q=15$, we obtain an expression that gives good accuracy for $6\leq q\leq24$.

We write
\begin{equation}\label{Tc_Taylor}
    \frac{\Tc^*(q)}{J} = \sum_{k=0}^\infty a_k (q-15)^k.
\end{equation}
\begin{figure}[H]
    \centering
    \includegraphics[scale=1]{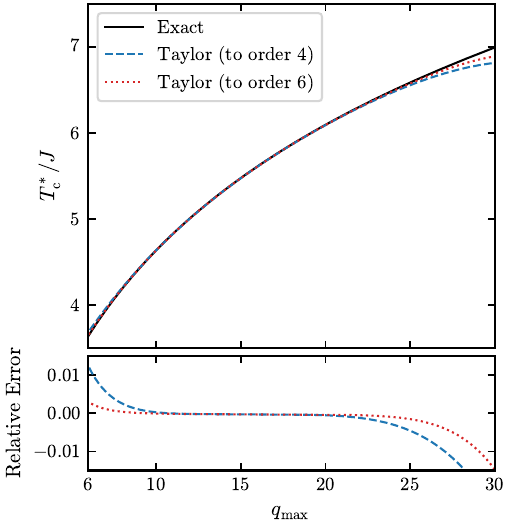}
    \caption{Taylor expansion of $\Tc^*(\qmax)$ around the point $\qmax=15$. The Taylor approximations can be computed via \cref{Tc_Taylor}, with the coefficients $a_k$ given in \cref{table_Taylor_coefficients}. Plotted here are Taylor approximations both to fourth order (dashed blue line) and to sixth order (dotted red line), while the solid black line shows the exact $\Tc^*$. The bottom panel shows the relative error associated with the two Taylor approximations. For $\qmax\leq30$, the relative error of the sixth-order approximation remains within approximately 1\%.}
    \label{fig_Taylor}
\end{figure}  
\noindent The first few coefficients $a_k$ are given in \cref{table_Taylor_coefficients}. Taking \cref{Tc_Taylor} and keeping only terms up to fourth order gives a relative error of less than $1\%$ for $7\leq q\leq27$. For more precise values, one can compute the sixth-order expansion for which the relative error is less than $10^{-3}$ for all $8\leq q\leq24$. These approximations are plotted in \cref{fig_Taylor}. If more precision is needed, one should implement \cref{tc_star_numerical}, truncating at a moderate value of $n$. For instance, choosing $n=30$ gives results with a relative error less than $10^{-4}$ for all $q\leq100$.

\end{appendix}

%

\end{document}